\documentclass[aps,pre,floats,floatfix,superscriptaddress, twocolumn]{revtex4-1}

\usepackage{latexsym,amsmath}
\usepackage{amsbsy}
\usepackage[pdftex]{graphicx}
\usepackage{amssymb}
\usepackage{epstopdf}
\usepackage[usenames]{color}
\usepackage{hyperref}
\usepackage{verbatim}
\usepackage{multirow,tabularx}

\newcommand{\av}[1]{\langle #1 \rangle}

\newcommand{\ZZZ}[2]{}

\newcommand{\FigPath}{./}

\begin{document}

\title{The interconnected wealth of nations:  \\ Shock propagation on global trade-investment multiplex networks }

\author{Michele Starnini}
\affiliation{Data Science Laboratory, ISI Foundation, Torino, Italy}
\author{Mari\'an Bogu\~n\'a}
\affiliation{Departament de F{\'\i}sica de la Mat\`eria Condensada, Universitat de Barcelona, Mart\'{\i} i Franqu\`es 1, 08028 Barcelona, Spain}
\affiliation{Universitat de Barcelona Institute of Complex Systems (UBICS), Universitat de Barcelona, Barcelona, Spain}
\author{M. \'Angeles Serrano}
\affiliation{Departament de F{\'\i}sica de la Mat\`eria Condensada, Universitat de Barcelona, Mart\'{\i} i Franqu\`es 1, 08028 Barcelona, Spain}
\affiliation{Universitat de Barcelona Institute of Complex Systems (UBICS), Universitat de Barcelona, Barcelona, Spain}
\affiliation{ICREA, Pg. Llu\'is Companys 23, E-08010 Barcelona, Spain}

\begin{abstract}
The increasing integration of world economies, which organize in complex multilayer networks of interactions, is one of the critical factors for the global propagation of economic crises. We adopt the network science approach to quantify shock propagation on the global trade-investment multiplex network. 
To this aim, we propose a model that couples a Susceptible-Infected-Recovered epidemic spreading dynamics, describing how economic distress propagates between connected countries, with an internal contagion mechanism, describing the spreading
of such economic distress  within a given country.
At the local level, we find that the interplay between trade and financial interactions
influences the vulnerabilities of countries to shocks.
 At the large scale, we find a simple linear relation between the relative magnitude of a shock in a country and its global impact on the whole economic system, albeit the strength of internal contagion is country-dependent and the inter-country propagation dynamics is non-linear. 
 Interestingly, this systemic impact can be predicted on the basis of intra-layer and inter-layer scale factors that we name network multipliers, that are independent of the magnitude of the initial shock. 
 Our model sets-up a quantitative framework to stress-test the robustness of individual countries and of the world economy to propagating crashes.
\end{abstract}

\date{\today}
\maketitle

\section*{Introduction}
\label{sec:intro}

The integrated nature of the international economy is the ultimate cause for the propagation of economic crisis at a global scale~\cite{RePEc:oxp:obooks:9780199251421}. A shock originated in a country, indeed, may spread to his economic partners through multiple flows,  captured by its balance of payments~\cite{RePEc:tcd:tcduee:20014}.
Shocks can have different origins, e.g. financial shocks can be caused by defaults of big financial institutions or sovereign debt crisis, while trade shocks may be triggered by barriers raised by governments, such as protective tariffs. 
The increasing {global interconnectedness of world economies}~\cite{RePEc:aea:aecrev:v:100:y:2010:i:2:p:388-92} calls for a modeling framework of shock propagation able to incorporate the full complexity of these interactions.

Network science~\cite{Newman2010} has established as the theoretical foundation that allows to quantify, model, and predict the behavior of spreading phenomena in complex systems, from information diffusion over social networks to epidemic contagion in living systems~\cite{Bakshy:2012:RSN:2187836.2187907, Pastor-Satorras:2014aa}. 
The study of international trade networks, such as the World Trade Web (WTW), has a long tradition in network science~\cite{PhysRevE.68.015101,Garlaschelli:2004aa,Serrano2007,Hidalgo482,Fagiolo:2009aa,serrano:2010,De-Benedictis:aa,Garcia-Perez:2016aa}. {Within the framework of the spreading of economic crisis, it has been proved that networks effects can be substantial}~\cite{Lee:2011aa, RePEc:imf:imfwpa:15/149}.
Conversely, network tools have been increasingly employed to estimate systemic risk among financial institutions~\cite{Caccioli2018, NBERw18727, Battiston10031, RePEc:hal:journl:hal-00912018, Battiston:2012aa, RePEc:ssa:lemwps:2013/08}, e.g. by adopting threshold models to assess financial contagion over networks of banks~\cite{Espinosa_Vega_2011, Gai2401, doi:10.1080/14697688.2014.968356}. 
While most of these works assess financial stability by considering the failure of single institutions (e.g., banks), and specific propagation channels (e.g. interbank lending), fewer works considered global networks at the country level~\cite{MINOIU2013168, Joseph:2014aa} and addressed shock propagation over financial cross-border networks~\cite{RePEc:ecb:ecbwps:20091124, RePEc:imf:imfwpa:16/91, RePEc:imf:imfwpa:18/113}. 

However, international shocks in the real world spread through both trade and financial relations. Neglecting the interplay between these channels may lead to underestimate spillover effects.
In this paper, we address international shock propagation by taking into account 
{both trade and financial} international relations, represented as a multiplex network~\cite{Boccaletti2014} 
 reconstructed by using {yearly} data of bilateral trade and financial positions between countries,  coupled to a dynamics describing how economic distress spreads from one country to another.  
Our model allows to estimate both the vulnerability of a country to external shocks, and the systemic {impact} that a country poses to the whole economic system. Remarkably, we find that spillover effects due the interconnectedness of economic relations can be encoded into networks multipliers, which allow predicting the global impact of economic shocks.

\section*{Multiplex representation of trade-investment networks} 
\label{sec:multiplex}
We assume that a shock can be transmitted {through} two main channels, trade and investment, forming the different layers of a multiplex network, the global trade-investment (GTI) network. In GTI networks, countries are represented by nodes connected in {each} layer by weighted links, representing the corresponding economic interactions and their intensity. 
The trade ($T$) layer of a GTI multiplex network is reconstructed by means of bilateral trade data of goods exchanges. 
We used the United Nations Commodity Trade Statistics Database~\cite{comtrade} (analyzed for the first time as a complex network in~\cite{PhysRevE.68.015101}), as detailed in {the Methods section}. 
A directed link from {country} $i$ to {country} $j$ in layer $T$ represents the exports of goods from $i$ to $j$ in a given year, $x_{ij}$, {which is} equivalent to {the} imports of goods of $j$ from $i$, $m_{ji} \equiv x_{ij}$. The total exports of country $i$ are simply $X_i= \sum_j x_{ij}$, and its total imports are $M_i = \sum_j x_{ji} = \sum_j m_{ij} $.
Transactions in goods account for the majority (generally over $70\%$) of the current account of a country, and thus they can be considered as a good proxy of the strength of its trading interactions, see Supplementary Information (SI) for details. 

In the investment ($I$) layer,
we consider cross-border positions of portfolio securities between two countries, reported in Ref.~\cite{qjt012}, as a proxy of the strength of their financial interactions, see Methods and SI. 
The $I$ layer is thus reconstructed  
 such that a link directed from node $i$ to node $j$ represents the stock of portfolio assets owned by country $i$ and issued by country $j$ in a given year, $a_{ij}$, equivalent to a portfolio liability for $j$ to $i$, $l_{ji} \equiv a_{ij}$. The total stock of portfolio assets owned by $i$ in a given year is simply $A_i= \sum_j a_{ij}$, and its total portfolio liabilities reads $L_i = \sum_j a_{ji} = \sum_j l_{ij}$. 
Note that while the $T$ layer is formed by trade flows, links in the $I$ layer represent stock quantities. 
Finally, note that the following trivial relations hold
\begin{equation}
\label{eq:tot_W}
W_T = \sum_i X_i = \sum_i M_i  , \qquad W_I = \sum_i A_i = \sum_i L_i 
\end{equation}
where $W_T$ stands for the annual total value of traded goods, and $W_I$ for the annual total value of investment positions. 
By considering the data sets available in ~\cite{qjt012} (see Methods and SI), GTI multiplex networks can be reconstructed for each year between 2001 and 2008. In the rest of the paper, as an illustrative case we consider the GTI network corresponding to the year 2005 (whose topological properties are described in the SI), results for other years are similar. 

\begin{figure*}[tb]
  \begin{center}  
    \includegraphics[width=17cm,angle=0]{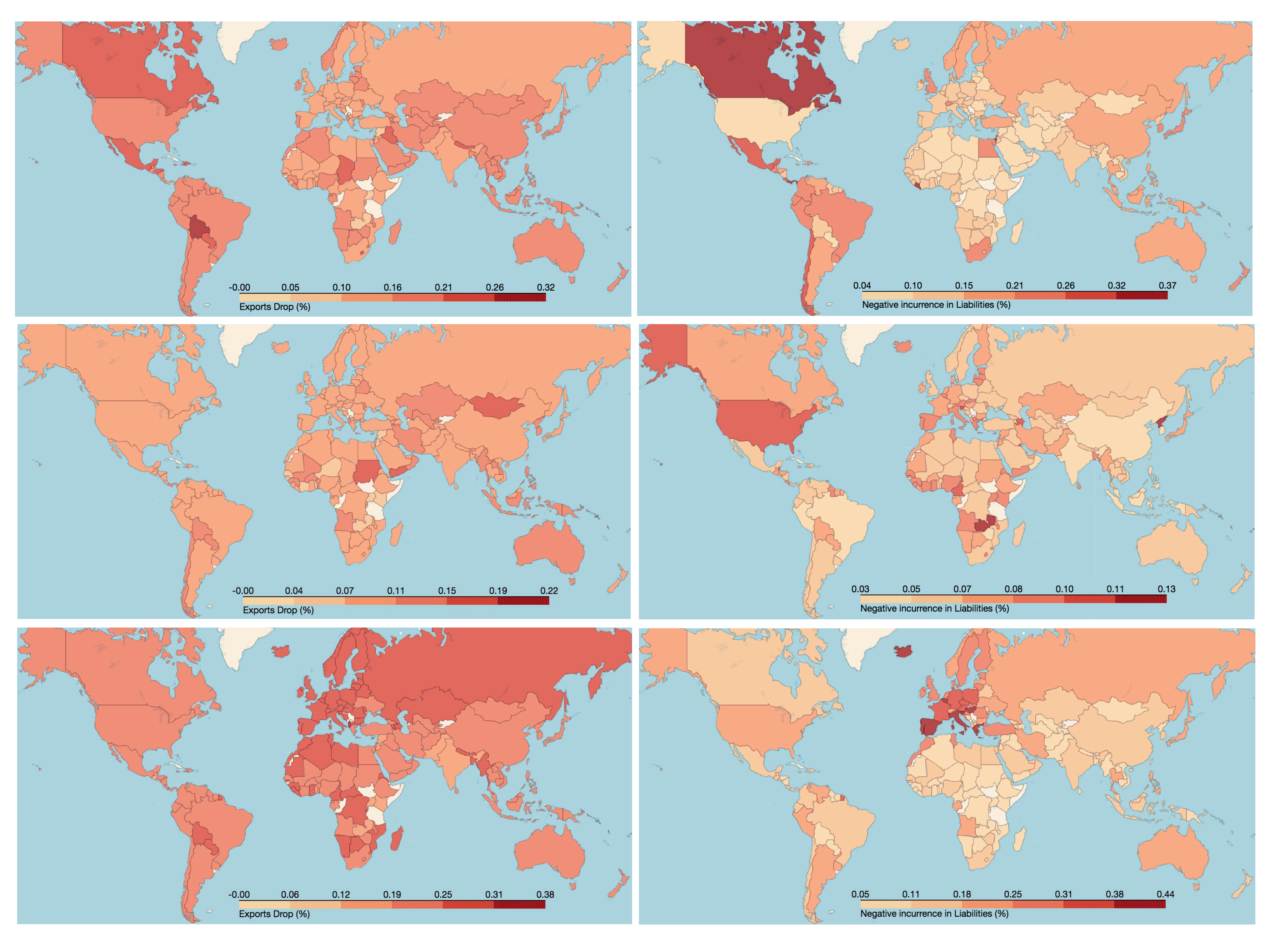}
      \end{center}
      \caption{ 
   Vulnerability of each country with respect to a shock originated in the United States (first row), China (second row), or in countries belonging to the EZ (third row), characterized by  $\alpha=-0.4$ and $\beta=-0.1$.   
      Colors indicate the VaR of exports, $VaR[\Delta X_i ]$ (left plots), and of incurrence in liabilities, $VaR[\Delta L_i]$ (right plots). 
         \label{fig:maps}}      
\end{figure*}

\section*{The shock propagation model} 
\label{sec:shock_prop}

A shock in an epicenter country may be driven by different domestic or exogenous factors, such as political instability, fiscal contraction, banking crisis, etc. 
Here we assume that shocks cause a drop on aggregate demand. 
This implies that the epicenter country may reduce its imports from other countries and/or its investment in financial assets issued by other countries. 
An initial shock in a country $i$ can be fully characterized by two parameters $\alpha$ and $\beta$, representing the initial variations in imports and foreign assets investment, $\delta M_i(t=0) = \alpha$ and $\delta A_i(t=0) = \beta$, {respectively}.
The notation $\delta Y_i(t) \equiv \frac{Y_i(t) - Y_i(t-1)} {Y_i(t-1)}$ stands for the relative variation of $Y_i(t)$ in time, $Y_i = \{X_i, M_i, A_i, L_i \}$,
where time $t$ is accounted by discrete time steps {in the shock propagation process,} $t=0,1, \ldots, t_{end}$. 
The distress is subsequently distributed from country $i$ to its partners, proportionally to the intensity of the  economic interactions of the corresponding channels. This implies negative variations in the exports and liabilities --$\delta X_j(t)$ and $\delta L_j(t)$-- of impacted neighbors $j$. On their turn, these variations may produce variations in imports, $\delta M_j(t)$, and assets acquisition, $\delta A_j(t)$, of each country $j$, that will be again distributed proportionally to
their neighbors  in the GTI multiplex, and so on. 
Therefore, the model's behavior is defined by a coexistence of two coupled but different dynamics:
i) the external propagation of the shock from distressed to {connected} countries, and ii) the internal {contagion} of the shock within distressed countries.

The inter-country contagion is modulated by a Susceptible-Infected-Recovered (SIR) spreading dynamics \cite{Pastor-Satorras:2014aa} on the GTI network, {which properly accounts for} reverberation and second-order effects. 
Each country is classified in three mutually exclusive states: susceptible to receive the shock for the first time, infected if it has accumulated distress and is able to propagate it, or recovered (inactive) when it can receive distress from its partners but cannot propagate it. Initially, all countries are in the susceptible state, except for the epicenter country, which is infected. 
Infected nodes spread their distress to all neighbors (regardless of their status), 
and turn inactive immediately after.
Susceptible countries reached by the propagation become infected.
The shock propagation continues until all infected countries have spread their accumulated distress, and the infected state disappears from the system. 
 Then, the SIR contagion dynamics is repeated several times, {each time} setting as initial {variations} the distress accumulated by inactive nodes in the last round, until the system reaches a final steady state (see SI for a concrete example).   

The internal {contagion} of the shock within distressed countries can be modeled by relating the variation of imports and assets acquisition of a country at time $t$ to the variation in its exports and liabilities incurrence in the short term, such as $\delta M_t = f(\delta X_t, \delta L_t)$ and $\delta A_t = f(\delta X_t, \delta L_t)$ (we omit the dependency in the country label for brevity).
By considering balance of payments constraints, indeed, the capacity of a country to pay for its import and to acquire foreign assets may depend on its ability to generate sufficient revenues from exports and financial liabilities.  We thus assume that a country's revenues from exports and financial liabilities can be viewed as a budget constraint on its capacity to import and acquire foreign assets. 
 We also assume that the exchange rate and prices do not adjust quickly and we neglect possible policy action aimed at counteract the shock effects.
 
 Therefore, the dependency of imports and financial assets on revenues from exports and liabilities can be viewed as a simple elasticity relation, and learned from the data. 
  To this aim, for each country we consider a multivariate linear regression model representing the correlations between the quantities $\delta X_t$, $\delta M_t$, $\delta A_t$, and $\delta L_t$. 
The elasticity relations can be described by the following equations: 
\begin{eqnarray}
\delta M_t & = & c_M + c_{MX}\, \delta X_t + c_{ML}\, \delta L_t +\epsilon_M \nonumber \\
\delta A_t & = & c_A + c_{AX}\,\,  \delta X_t + c_{AL} \,  \,   \delta L_t +\epsilon_A,
\label{eq:multilinear_model}
\end{eqnarray}
where terms $c_M$ and $c_A$ represent the variation trend of terms $dM_t$ and $dA_t$, respectively,
the coefficients  $c_{MX}$, $c_{AL}$, $c_{ML}$, and $c_{AX}$ encode the correlations between $(dX_t,dL_t)$ and $(dM_t,dA_t)$, while  $\epsilon_M$ and $\epsilon_A$ account for Gaussian noise, 
with zero average $\langle  \epsilon_M \rangle = \langle  \epsilon_A \rangle = 0$ and variance $\langle \epsilon_M^2 \rangle = \sigma_{\epsilon_M}^2$, $\langle \epsilon_A^2 \rangle = \sigma_{\epsilon_A}^2$.
It is important to remark that Eqs.~\ref{eq:multilinear_model} are treated as simultaneous equations, by incorporating the possible correlations between all variables, see Methods and SI.
Coefficients  $c_{MX}$, $c_{ML}$, $c_{AX}$, and $c_{AL}$ thus play the role of \textit{internal pass-through coefficients}, since they describe how the variations of imports and asset of a country depend on the variations of its exports and liabilities. 
Countries with 
internal pass-through coefficients smaller/larger than one will reduce/{increase} the impact of the shock to their commercial or financial partners, acting thus as 
absorbers/amplifiers. For instance, oil exporters play the role of shock blocker, having small internal pass-through coefficients. See Methods and SI for a detailed description of internal pass-through coefficients and their estimation. 

\section*{Vulnerability of countries to propagating shocks}
\label{sec:vulnerability}

The shock propagation model 
allows us to assess the impact of demand shocks in one or more countries on the rest of the world,  when the shock spreads from one country to another through international macroeconomics networks like the GTI multiplex. 
The impact on a country $i$ produced by a shock originated in an epicenter country $E$, with parameters $(\alpha,\beta)$, can be quantified by considering the relative variations $\Delta Y_i(\alpha, \beta, E)$ of each macroeconomic variable of country $i$, $Y_i = \{ X_i, M_i, A_i, L_i \}$, measured 
at the end of the system's evolution (once the shock has been totally absorbed by the entire system)
with respect to its initial value, that is,
 \begin{equation}
 \label{eq:vulnerability}
V_i(Y|{\alpha, \beta, E})\equiv\Delta Y_i({\alpha, \beta, E}) = \frac{Y_i(t_{end}) - Y_i(t_0)}{Y_i(t_0)}.
\end{equation}
The quantity $V_i(Y|{\alpha, \beta, E})$ gives a measure of the {vulnerability} of country $i$ to a shock originated in country $E$. 
This magnitude can be very heterogeneous across different countries and, even for the nearest neighbors of the epicenter country $E$, it incorporates systemic effects beyond direct bilateral economic interactions.
By running several numerical simulations of the model with the same initial conditions $(\alpha, \beta, E)$, one can obtain probability distributions for the quantities  $V_i(Y|\alpha, \beta, E)$, and consequently 
the average $\av{V_i(Y|\alpha, \beta, E)}$ and value at risk $VaR[ V_i(Y|\alpha, \beta, E) ]$, as measures of the expected variability and the risk of loss.

Fig.~\ref{fig:maps} shows the heterogeneity of the vulnerability $V_i(Y|\alpha, \beta, E)$ across the world, for a shock characterized by parameters $(\alpha=-0.1,\beta=-0.4)$ 
and three different epicenter countries: the United States, China, and the Eurozone (EZ). 
We plot both the impacts on trade and investment, by coloring countries according to their VaR of exports, $VaR[V_i(X|-0.1,-0.4, E) ]$ (left plots), and incurrence in liabilities, $VaR[V_i(L|-0.1,-0.4, E) ]$ (right plots). 
One can see that American countries are more vulnerable to a shock originated in the United States (first row), with respect to both trade (the exports of Mexico, Canada and other South American countries may drop more than $25\%$) and investment. 
Shocks in China (second row) have a considerable lower impact on the rest of the world, especially with respect to investments. Australia, some African and South American countries may be forced to reduce their exports (probably raw materials) up to $20\%$, while the United States shows one of the largest reduction of foreign investment, around $10\%$. 
The economic impact on trade of a shock involving all {Eurozone} countries (third row) is homogeneously distributed to the rest of the world, with a general reduction of exports around $30\%$, while the financial impact is much more heterogeneous: most vulnerable countries are Southern European countries, forced to reduce their liabilities by more than $40\%$, probably due to sovereign debt exposures. The average vulnerabilities $\av{V_i(X|-0.1,-0.4, E) }$ and $\av{ V_i(L|-0.1,-0.4, E) }$ show qualitatively similar behaviors, see SI.
 The model can thus be used to rank the vulnerabilities of different countries with respect to economic shocks, depending on its epicenter and magnitude.  
 
\begin{figure}[tb]
  \begin{center}  
    \includegraphics[width=8.5cm,angle=0]{\FigPath/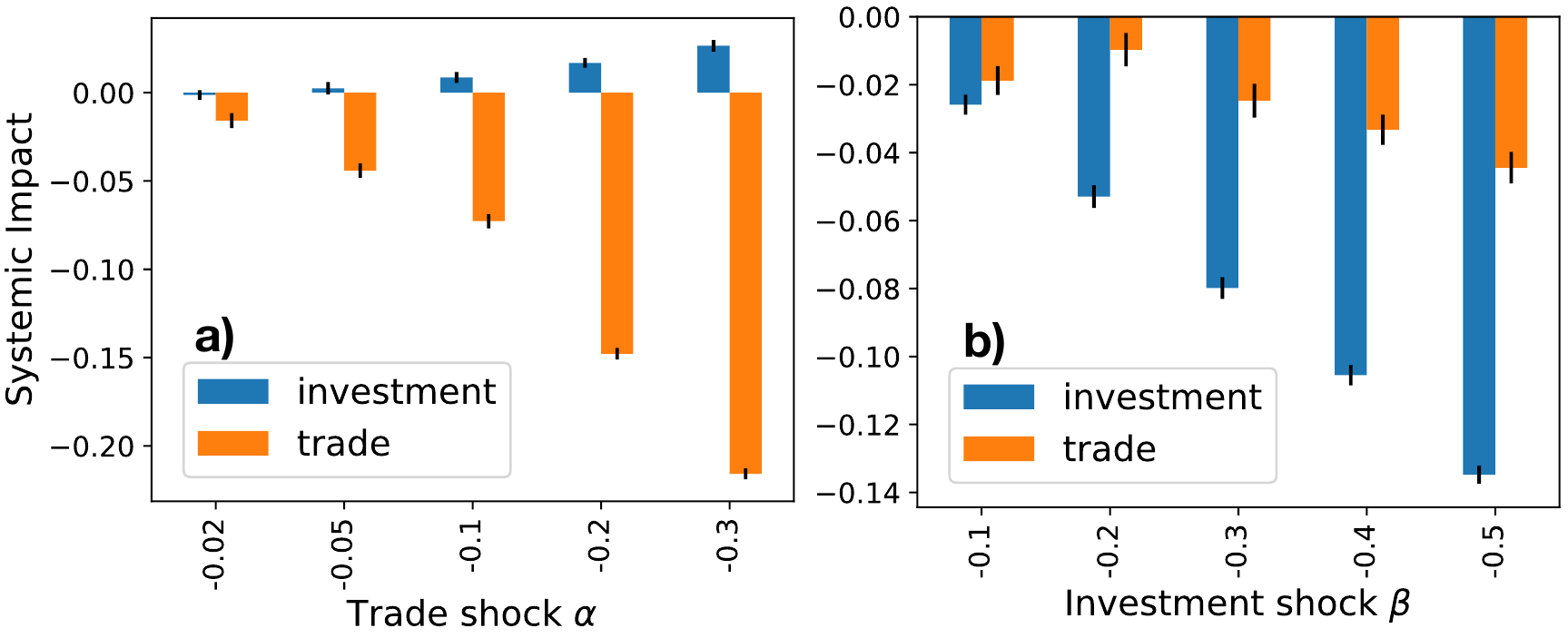}    
      \end{center}
      \caption{ 
       Systemic impact on trade, $\mathcal{S}_i^T(\alpha,\beta)$, and investment, $\mathcal{S}_i^I(\alpha,\beta)$, of a shock originated in the United States. 
 Different combinations of values $(\alpha,\beta)$ are considered: the initial shock can be originated in the trade layer (plot a), $\alpha=0$, $\beta < 0$), or trade layer (plot b), $\beta=0$, $\alpha < 0$). 
 Error bars represent the standard error of the mean over 100 runs. 
   A financial shock reducing by 40\% the foreign assets demand in a single, large country such as the United States is expected to reduce the total value of financial securities by 11\%, but also the total traded goods by 4\%. 
     \label{fig:interplay}}      
\end{figure}

\section*{Quantifying systemic impact of epicenter countries}
\label{sec:sys_impact}

Beyond {country} vulnerabilities, our model allows us to measure the potential risk that each country poses for the international economic system as a whole. 
The global impact of a shock in a given country across the GTI multiplex can be quantified by defining its \textit{systemic impact} $ \mathcal{S}_i(\alpha,\beta)$, as the total economic value that is {affected} by a shock originated in country $i$ with parameters $(\alpha,\beta)$. 
The systemic impact is expected to depend crucially on the propagation of the shock from financial to {trade layer}, and vice versa. 
These spillover effects between layers can be addressed by considering separately the impacts on trade and investment. One can define the systemic impact on trade, $ \mathcal{S}_i^T(\alpha,\beta)$, and investment $ \mathcal{S}_i^I(\alpha,\beta)$, as the {affected} value of traded goods and financial securities expressed as a fraction of the global value of traded goods $W_T$ and financial securities $W_I$, respectively,  that is 
 \begin{eqnarray}
  \mathcal{S}_i^T(\alpha,\beta) & = & \frac{\sum_j \av{ \Delta X_j(\alpha, \beta, i)}}{W_T}  = \frac{\sum_j \av{  \Delta M_j(\alpha, \beta, i}}{W_T}, \nonumber \\
  \mathcal{S}_i^I(\alpha,\beta) & = & \frac{\sum_j \av{  \Delta L_j(\alpha, \beta, i)}}{W_I} = \frac{\sum_j \av{  \Delta  A_j(\alpha, \beta, i)}}{W_I}.
\end{eqnarray} 
Note that the second equality holds because of \eqref{eq:tot_W}. 

Fig.~\ref{fig:interplay} shows the systemic impact on trade, $\mathcal{S}_i^T(\alpha,\beta)$, and investment, $\mathcal{S}_i^I(\alpha,\beta)$, of a shock originated only in the financial layer ($\alpha = 0$, Fig.~\ref{fig:interplay}a), or trade layer ($\beta = 0$, Fig. \ref{fig:interplay}b) of the GTI multiplex networks, with the United States as epicenter country. 
As expected, the larger the initial distress, represented by parameters $(\alpha, \beta)$, the larger the systemic impact on the rest of the world.
Intriguingly, even if the initial shock only involves one layer, the economic distress spreads from {the financial to the trade layer}, and viceversa. 

Different countries exhibit different magnitudes of systemic impact on trade or investment, that can be taken as a measure of their {relevance for the stability of the GTI multiplex.}
The systemic impact of a country $i$, indeed, is expected to depend on the economic value of the initial shock $\mathcal{I}_i$, determined simply as {$\mathcal{I}_i = (\alpha M_i + \beta A_i)/(W_I + W_T)$}. 
Fig.~\ref{fig:residuals}a shows the  systemic impact  on trade $\mathcal{S}_i^T$, and investment $\mathcal{S}_i^I$, as a function of the value of the initial shock $\mathcal{I}_i$, characterized by $\alpha = \beta = -0.2$, for countries belonging to the $G_{20}$ group. 
 Surprisingly, we found that the systemic impacts of these countries on global trade ($\ell=T$) or investment ($\ell=I$) are well fitted by linear regressions, whose coefficients $\gamma_{\ell}( \alpha, \beta)$ represent {scale} factors for the initial shock. 
This implies that, at least for big economies, the systemic impact of a country $i$ can be described simply as $\mathcal{S}_i^{\ell}( \alpha, \beta) \simeq \gamma_{\ell}( \alpha, \beta) \mathcal{I}_i$, where $\gamma_{\ell}( \alpha, \beta)$ encodes the {sensitivity of the GTI multiplex to the shock}. 
The larger the coefficients $\gamma_{\ell}( \alpha, \beta)$, the larger the shock amplification. 
Notice that these coefficients depend on the initial shock but are country-independent.
Even if the elasticity relations \eqref{eq:multilinear_model}, determining the internal contagion within countries, are linear, the pass-through coefficients are quite heterogeneous across countries (see SI), and the inter-country propagation phase modeled by the SIR dynamics introduces highly non-linear effects. 

Furthermore, it is interesting to consider the regression residuals of different countries. For each country $i$, one can define the deviations of the systemic impact of each country from the expected value obtained by the fitting function, as
\begin{equation}
 \mathcal{D}_i^{\ell}(\alpha, \beta) =  \gamma_{\ell}( \alpha, \beta) \mathcal{I}_i - \mathcal{S}_i^{\ell}(\alpha, \beta).
 \end{equation} 
The trade (financial) deviation $\mathcal{D}_i^{T}$ ($\mathcal{D}_i^{I}$) of a country $i$ can be positive, if its systemic impact on trade (investment) is smaller than the fitted value, or negative, if $\mathcal{S}_i^{T}$ ($\mathcal{S}_i^{I}$) is larger than what expected by considering the magnitude of the initial shock $\mathcal{I}_i$.
Fig.~\ref{fig:residuals}b shows the trade and financial deviations, $\mathcal{D}_i^T$ and $\mathcal{D}_i^I$, respectively, of the systemic impact of each country $i$ belonging to the $G_{20}$ group. 
These deviations are affected by both the statistical error on the systemic impact and the uncertainty of the fitting function, and thus few countries show statistically significant values of $ D_i^{\ell}(\alpha, \beta)$. 
However, one can see that countries having a significant, positive deviation on trade, generally show a significant, negative deviation on investment, and viceversa. China and Germany, for instance, have a larger systemic impact on trade and a smaller impact on investment than expected, while the United Kingdom and Japan show a considerably smaller impact on trade and a larger impact on investment.
Even though $ D_i^{\ell}(\alpha, \beta)$ are expected to depend on the magnitude of the initial shock, these countries presenting significant values of the deviations have qualitatively similar behavior regardless the value of $(\alpha, \beta)$, as shown in the SM. 
It is worth to note that it is not possible to verify the linear scaling between initial shock and systemic impact, and consequently its deviations, for small economies, due to large uncertainties over the impact of these countries. 

\begin{figure}[tb]
  \begin{center}  
    \includegraphics[width=8.5cm,angle=0]{\FigPath/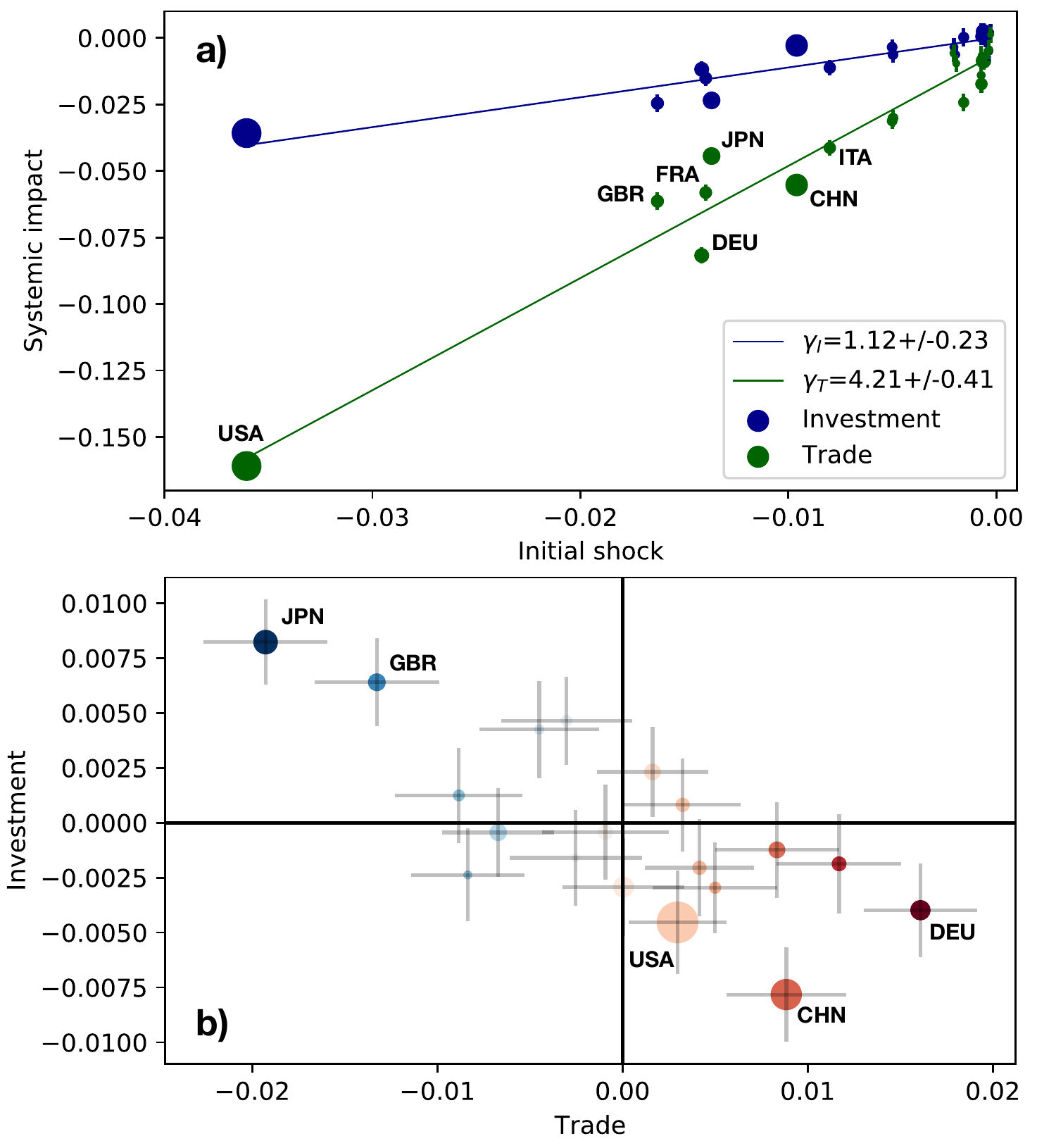}
      \end{center}
      \caption{ 
   a) Systemic impact on global trade $\mathcal{S}_i^T$ and investment $\mathcal{S}_i^I$, as a function of the magnitude of the initial shock $\mathcal{I}_i/ (W_I + W_T)$. b) Trade ($\mathcal{D}_i^T$, x-axis) versus  financial ($\mathcal{D}_i^I$, y-axis) deviations, as obtained by plot a). 
   The initial shock is characterized by $\alpha = \beta = -0.2$ (different values in the SI), countries belonging to the $G_{20}$ group are shown. Error bars represent the standard error of the mean for  $\mathcal{S}_i$. Regression coefficients $\gamma_{\ell}$ are plotted with $95\%$ CI. Size of dots is proportional to countries' GDP.   \label{fig:residuals} }     
\end{figure}

\section*{Network multipliers predict systemic impact}
\label{sec:net_multipliers}

\begin{figure}[tb]
  \begin{center}  
      \includegraphics[width=8.3cm,angle=0]{\FigPath/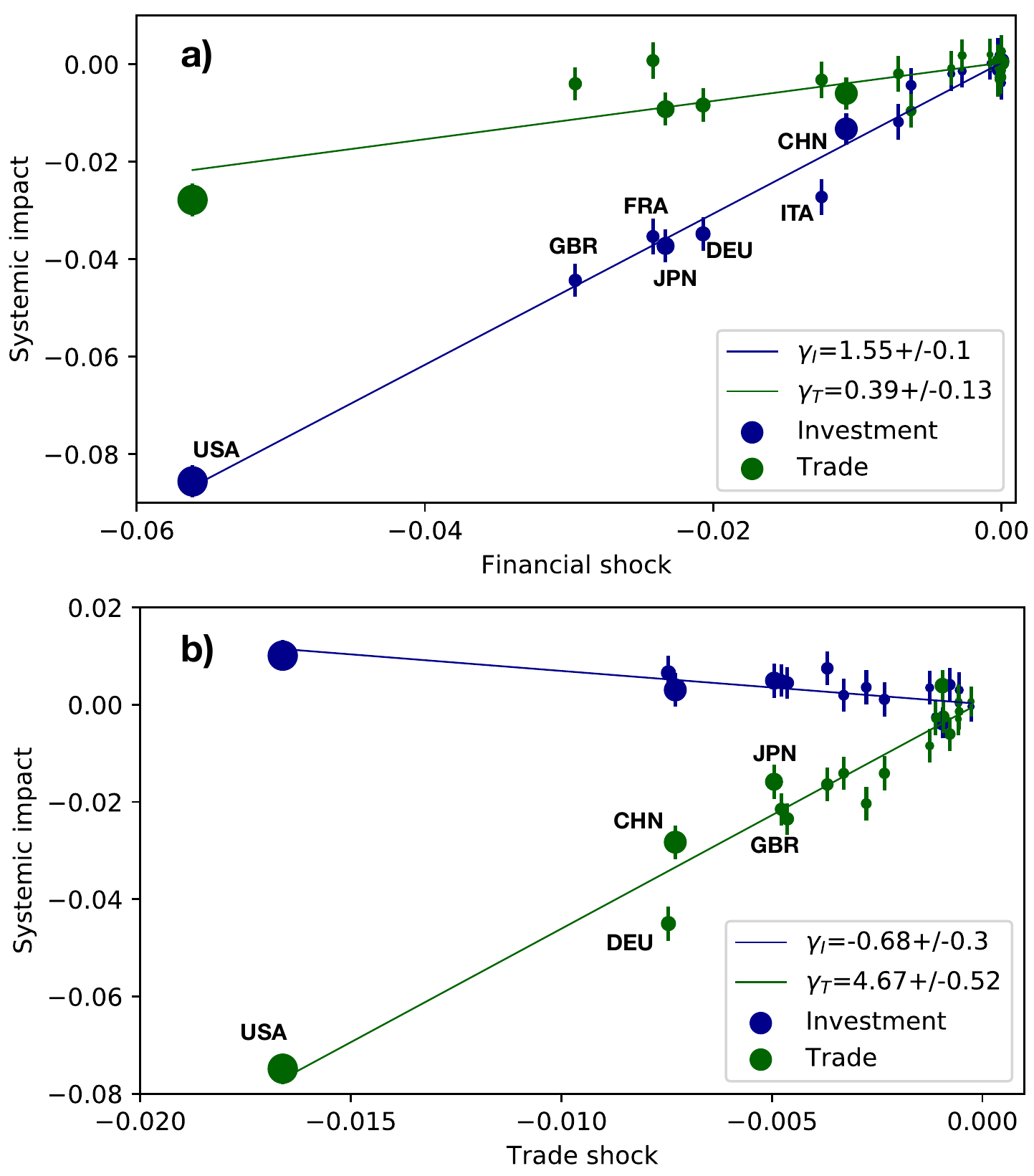}
 \end{center}
      \caption{ 
     Systemic impact on global trade $\mathcal{S}_i^T$ and investment $\mathcal{S}_i^I$, as a function of the an initial shock $\mathcal{I}_i^{\ell}/W_{\ell}$ originated only in the investment (plot a), $\alpha=0$, $\beta=-0.3$) or trade (plot b), $\alpha=-0.1$, $\beta=0$, right) layer, for countries belonging to the $G_{20}$ group. Different values of $(\alpha, \beta) $ are shown in the SI. Error bars represent the standard error of the mean for  $\mathcal{S}_i$. Regression coefficients $\gamma_{\ell' \rightarrow \ell}$ are plotted with $95\%$ CI. Size of dots is proportional to countries' GDP.  
       \label{fig:gen_imp_in_shock}}     
\end{figure}

The value of the coefficients $ \gamma_{\ell}( \alpha, \beta)$ depends on the parameters $( \alpha, \beta)$ characterizing the initial shock (see SI). One can understand this dependency by considering separately shocks originated only in one layer, investment or trade, of the GTI multiplex. 
Fig.~\ref{fig:gen_imp_in_shock} shows that, also in the case of a {exclusively} financial ($\alpha=0.1$, Fig.~\ref{fig:gen_imp_in_shock}a) or  {exclusively} trade ($\beta=0.3$, Fig.~\ref{fig:gen_imp_in_shock}b) shock, the systemic impacts  $\mathcal{S}_i^T$ and $\mathcal{S}_i^I$ are well fitted by linear regressions. 
However, the regression coefficients do not strongly depend on the magnitude of the initial shock, being remarkably similar for different values of $(\alpha, \beta)$, see SM. 
Therefore,  we name the scale factors $\gamma_{\ell' \rightarrow \ell}$ as \textit{intra- and inter-layer network multipliers}, as they gauge the network effects of shock propagation from layer $\ell$ to layer $\ell'$ on GTI networks,
\begin{equation}
\mathcal{S}_i^{\ell}(\alpha,\beta) \simeq \gamma_{\ell' \rightarrow \ell} \,  \mathcal{I}_i^{\ell'},
   \end{equation}    
{where $\mathcal{I}_i^{T}=\alpha M_i/W_T$ and $\mathcal{I}_i^{I}=\beta A_i/W_I$.}

Fig.~\ref{fig:gen_imp_in_shock}, S9 and S10 show that a financial shock has an \textit{intra-layer} {network multiplier} $\gamma_{I \rightarrow I} \simeq 1.5 \pm 0.1$, and a \textit{inter-layer} {network multiplier} $\gamma_{I \rightarrow T} \simeq 0.3 \pm 0.15$. 
As expected, the {network multiplier for the systemic impact on the investment layer}, $\gamma_{I \rightarrow I}$,  is much larger than the 
one for the trade layer, $\gamma_{I \rightarrow T}$. 
Conversely, a trade shock shows an intra-layer network multiplier of $\gamma_{T \rightarrow T} \simeq 4.5 \pm 0.5$, and an inter-layer {network multiplier of} $\gamma_{T \rightarrow I} \simeq -0.6 \pm 0.3$. 
It is interesting to note that the {network multiplier giving the systemic impact on trade for a trade shock} $\gamma_{T \rightarrow T}$  is much bigger than the {network multiplier giving the systemic impact on investment for an investment shock} $\gamma_{I \rightarrow I}$, meaning that the {intra-layer network effects are stronger for trade shocks than} for financial shocks. 
The {network multiplier} $\gamma_{T \rightarrow I}$ is negative, indicating that trade shocks can produce an increase in the incurrence in liabilities, probably to compensate the revenue reduction from exports.

\begin{figure}[tb]
  \begin{center}  
      \includegraphics[width=8.5cm,angle=0]{\FigPath/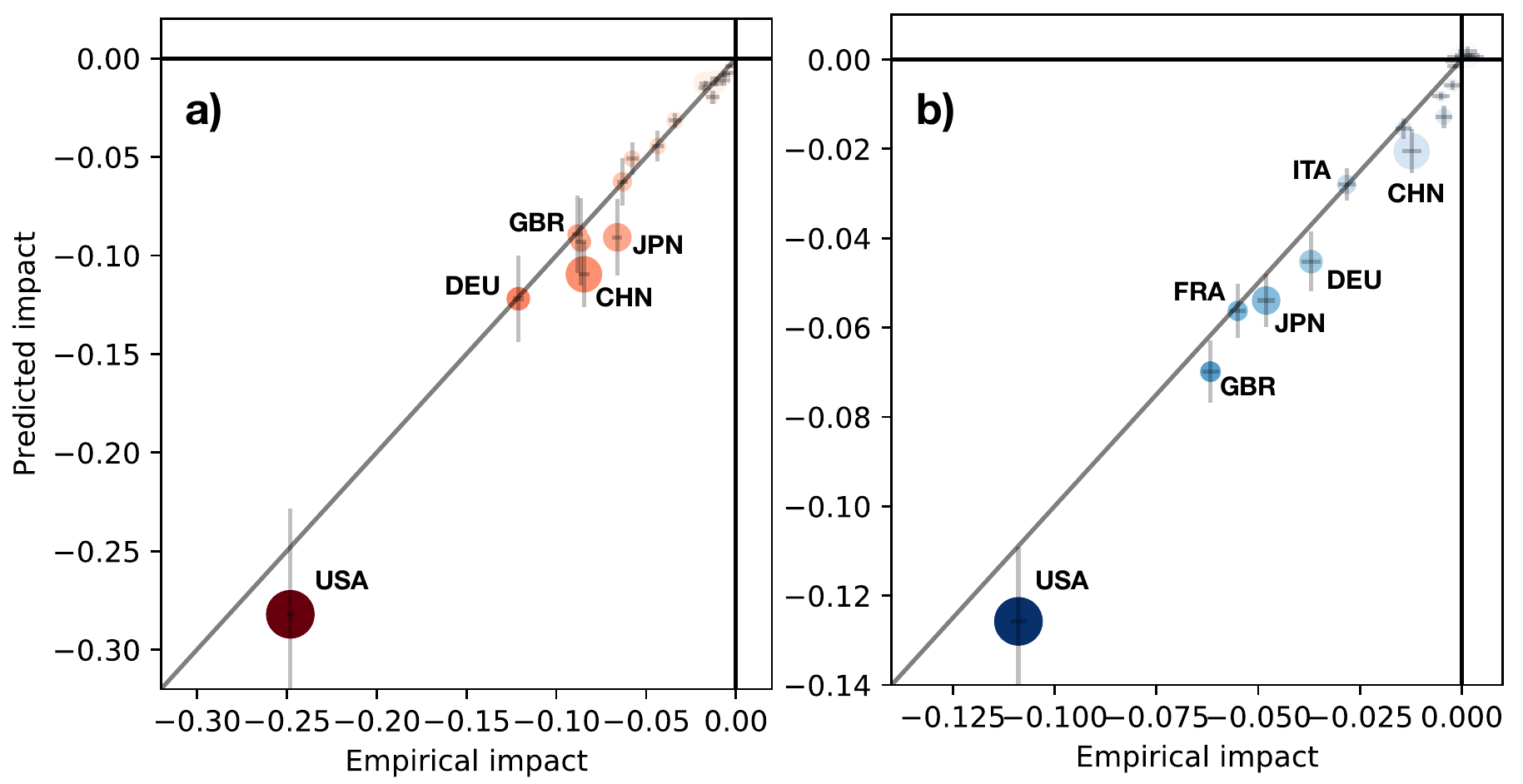}
\end{center}
      \caption{ Expected versus actual systemic impact on trade (a) and investment (b) of each country $i$ belonging to the $G_{20}$ group, originated by an initial shock with $\alpha = -0.3, \beta = -0.5$.  
      The size of dots is proportional to their GDP, color proportional to $\mathcal{S}_i^{\ell}$ (red for $\ell=T$, blue for $\ell=I$). Uncertainties are represented by grey crosses.  \label{fig:pred}   }
\end{figure}

  The {network multipliers} $\gamma_{\ell \rightarrow \ell'}$ may be used to predict the systemic impact of a country {hurt by a combined or single-layer shock}, given its relative magnitude {in each layer}. 
  If we assume that the systemic impact generated by an initial shock in both financial and trade layers, characterized by $(\alpha, \beta)$, is comparable to the sum of the systemic impacts of a trade shock with $\alpha$, and a financial shock with $\beta$, 
  then one can estimate the expected impact as
  \begin{equation}
\left( \begin{array}{c} \mathcal{S}_i^{T}(\alpha, \beta)  \\ \mathcal{S}_i^{I}(\alpha, \beta) \end{array} \right) \simeq
 \begin{pmatrix} \gamma_{T \rightarrow T} & \gamma_{I \rightarrow T} \\ \gamma_{T \rightarrow I} & \gamma_{I \rightarrow I} \end{pmatrix}  \left( \begin{array}{c} \mathcal{I}_i^{T}(\alpha)  \\ \mathcal{I}_i^{I}( \beta)  \end{array} \right)
 \label{eq:pred}
  \end{equation}
Fig.~\ref{fig:pred} shows a comparison between the expected impact on trade (Figure \ref{fig:pred}a) and investment (Figure \ref{fig:pred}b), as derived from \eqref{eq:pred}, and the actual systemic impact originated by an initial shock with $\alpha = -0.3, \beta = -0.5$. 
One can see that, by taking into accounts the statistical error on the systemic impact and the uncertainty on the {network multipliers} $\gamma_{\ell \rightarrow \ell'}$, expected and actual impacts are {actually very close. Thus,} \eqref{eq:pred} allows to predict the systemic impact of a country, given the initial shock (see SI for different values of $\alpha, \beta$).

Finally, Figs.~\ref{fig:pred} and S11 show clearly that the systemic impact of a country does not only depend on its GDP, and may be significantly different for trade or investment. The country with the largest systemic impact on the rest of the world is by far the United States, with respect to both trade and investment. However, the next countries with the largest impact on investment are the United Kingdom and Japan, while  Germany and China have the next largest impact on trade. Note that China, with the second largest GPD, has an expected impact on global investment ten times smaller than the United States.

\section*{Discussion}

{Estimating the global effects of economic crises remains a major challenge to be solved to advance in their prevention and control. We have proved here that a modeling strategy combining a multilayer network approach with inter-country and intra-country contagion dynamics is useful to stress-test the robustness of individual countries and of the world economy to propagating shocks.} 
Our model allows to estimate the different vulnerability of countries, that can act as absorbers or amplifiers.
{At large scale, the simple linear relation between the relative magnitude of a shock at the country level and its impact on the global system is surprising, since the strength of internal contagion is country dependent and the inter-country propagation dynamics is non-linear. Interestingly, this systemic impact can be predicted on the basis of intra-layer and inter-layer network multipliers, that are independent of the magnitude of the initial shock.}

It is important to remark that our modeling framework has several well-known limitations.  
{Financial data 
are still scarce for specific economies and often show inconsistency, caused for instance by tax havens~\cite{qjt012}. The missing information should then be estimated 
 at the risk of adding noise coming from the estimation methodologies.} 
{On the other hand, o}ur stress-test model represents a solid but first step towards a more sophisticated quantitative framework. For instance,
we did not take into account the possible variation of optimal decision rules of economic agents as a response to policy change~\cite{lucas_critique} and, 
in order to minimize the number of assumptions, the complexity of the economic structure of a country is neglected.
Furthermore, there might be several sources of endogeneity in determining {internal pass-through } coefficients through  \eqref{eq:multilinear_model}, such as omitted variables (e.g., GPD variation), which may lead to biased estimations { of the parameters}~\cite{Wooldridge2008}. 

Nevertheless, our approach aims at overcoming more serious limitations in the current modeling approach of global shock propagation, mostly based on threshold models, in which a node's failure triggers cascade dynamics. {Even} if the complete collapse of a financial institution has been empirically observed several times, the default of one or more countries, implying the complete stop of trade and financial flows, seems a very unrealistic assumption.
Finally, the linearity assumed in the inter-country phase of the shock propagation (another limitation, yet common in standard econometrics) is at least partially compensated by the non-linearity of the intra-country phase, originated from the repeated SIR dynamics.

One natural spinoff of our work would be the analysis of the evolution of the GTI multiplex network topology, intra-layer pass-through coefficients, and network multipliers to compare pre- and post-crisis scenarios. In the long run, we hope that our network-based macroeconomic approach to the propagation of shocks could be enriched and contribute to the detection of early warning signals, as well as suggest regulatory strategies to prevent the social, economic, and ecological cost of crises.

\section*{Methods and Materials}

Here we describe the empirical data used in the paper, available through motivated request to the authors, and the estimation of the internal pass-through coefficients of the shock propagation model. 

\subsection*{Empirical data}

Our work relies on different data sources, described in details in the SI, and summarized here. 
The investment layer of the GTI network is reconstructed by using the bilateral matrix of cross-border financial position between countries, as reported in Ref.~\cite{qjt012}. 
Bilateral data disclosing financial exposures are scarce. However, the Coordinated Portfolio Investment Survey (CPIS) annually conducted by the International Monetary Fund (IMF) reports data of cross-border positions of portfolio securities between countries.
Portfolio securities represent the largest fraction of cross-border investment positions, that include also direct investments and banking sector positions~\cite{RePEc:imf:imfwpa:18/113}.  
 We consider cross-border portfolio investment positions between two countries as a proxy of the strength of their financial interactions.
Note that, since CPIS data are biased because of offshore tax havens, here we considered the data sets compiled in Ref.~\cite{qjt012}, which completed CPIS data, as detailed in the SI.
 The trade layer of the GTI multiplex network is reconstructed by using the United Nations Commodity Trade Statistics Database~\cite{comtrade}, also used and described in Ref.~\cite{Garcia-Perez:2016aa}.  
The multivariate regression model, described by~\eqref{eq:multilinear_model}, is informed by the time series of exports, imports, incurrence of liabilities, and acquisition of assets, as recorded by the IMF. We considered yearly data, from 1980 to 2015. We exclude global recession periods from the time series, i.e. years 1982, 1991, and 2009~\cite{weo}. 

\subsection*{Estimation of internal pass-through coefficients}

We estimate trend terms, internal pass-through coefficients, and noise terms in \eqref{eq:multilinear_model} for each country by calculating variances and co-variances of the four time series $\{ dX_t, dM_t, dA_t, dL_t \}$, extracted from annual data recorded by the IMF
, as described in SI.
Some observations are in order. First, the model assumes that there are no lags between exports/liabilities revenues and imports/assets payments. 
Second, one can de-trend the relations described by \eqref{eq:multilinear_model} by setting trend terms equal to zero in the shock propagation dynamics, $c_A = c_M = 0$. 
Finally, note that correlations between terms $dM_t$ and $dA_t$ are directly related to the correlation between noises $\epsilon_1$ and $\epsilon_2$, as $\langle dM_t dA_t \rangle = \sigma_{\epsilon_1 \epsilon_2}^2$, see SI. 

\section*{Acknowledgements}

{We acknowledge support from {a James S. McDonnell Foundation Scholar Award in Complex Systems}; the ICREA Academia prize, funded by the Generalitat de Catalunya; Ministerio de Econom\'{\i}a y Competitividad of Spain project no.~FIS2016-76830-C2-2-P (AEI/FEDER, UE).}

\bibliographystyle{apsrev4-1}

\bibliography{Mende.bib}

\clearpage

\begin{widetext}

\section*{Supplementary Information}

\section{Empirical data sets description}
\label{sec:description-data}

In this section, we describe the empirical data sets used in the paper. Our work relies on the following data sources.

\begin{itemize}

\item The \textbf{Coordinated Portfolio Investment Survey (CPIS)} reports bilateral data on cross border portfolio investments between pairs of countries. The survey is conducted annually by the International Monetary Fund (IMF), started in 2001, and it distinguishes between equity and debt securities.  Participation in the survey by countries is voluntary.  The survey is conducted by asking a creditor country $i$ for its cross-border assets $a_{ij}$, issued by a debtor country $j$. The asset $a_{ij}$ is equivalent to a liability $l_{ji}$, issued by country $j$, owned by country $i$. The sum of assets owned by $i$ is $A_i = \sum_j a_{ij}$, and the sum of its liabilities reads $L_i = \sum_j l_{ij}$. Therefore, in the CPIS liabilities are derived for both reporting and non-reporting countries. If all countries reported in the CPIS, the resulting cross-border (portfolio) investment network would be a fully connected graph. For the year 2008, a total of 73 creditors (excluding important economies, such as China and oil exporters) reported on more than 200 debtors. Note that CPIS reports financial data only regarding portfolio investment, not including other components of the financial account (FA), such as directed investment (DI), financial derivatives (FD) or other investment (OI). 

The Cross-Border Investment Network (CBIN) is reconstructed from the bilateral matrix of cross-border financial position between countries, with data provided by the CPIS.
Since the CPIS data are incomplete and present problem of internal incoherence,  underestimating the net foreign asset positions of rich countries because of offshore tax havens, here we considered the data sets compiled in Ref. \cite{doi:10.1093/qje/qjt012}, 
that completed CPIS data in order to recover internal coherence, for years from 2001 till 2008.
Note that even in this case, for some countries, mainly Luxemburg, Cayman Islands and Ireland, the liabilities-side reported are largely incorrect, due to massive unreported investment in the financial industry of these countries, that lately re-direct such investment toward other countries. 
Reserves held by central banks and International organization are aggregated together, represented under the label SEFER+SSIO. We exclude this node from the network. 

\item The World Trade Web (WTW) is reconstructed from the data set compiled in Ref. \cite{Garcia-Perez:2016aa}. 
In the WTW, each node represent a country, and a the weight $w_{ij}$ of the direct link from $i$ to $j$ represents the amount of exports (in 2006 dollars) from country $i$ to country $j$. Since we are interest in a multiplex representation combining the WTW with the CBIN, also in this case we consider years from 2001 to 2008.

\item The shock propagation model, described in the main text and in more details in Section~\ref{sec:detail_shock_prop}, is informed by the time series of exports, imports, incurrence of liabilities, and acquisition of assets. 
Such time series are reported by the IMF as aggregated data (i.e. a single country vs rest of the world) of the balance of payments (BOP) for most countries. We informed the shock propagation model by yearly data from 1980 to 2015, excluding global recession periods from the time series, i.e. years 1982, 1991, and 2009. Depending on the country and time period, different level of detail is available (e.g. for South Africa, 1992, it is available the $FA$, but not its single components, such as DI or PI). Unfortunately, data of the BOP recorded by the IMF are generally not broken down to single counterparts, that is, bilateral data (i.e. a country $i$ vs another country $j$) are not reported by the IMF. The only data source for bilateral financial data is the CPIS. On the contrary, bilateral data regarding trade in goods have been collected by different sources, e.g. COMtrade.

\end{itemize}

\newpage

\section{Topology of GTI multiplex networks}
\label{sec:network_topology}

In this section, we describe how we reconstruct global trade-investment (GTI) networks, and discuss their topological properties.
International economic transactions, summarized by the balance of payments (BOP), can be represented as a multiplex network: each node represents a country, and different accounts (such as current and financial account) are described by different layers. 
Since we are interested in shock propagation, we reconstruct a network of ``vulnerabilities'' between countries, in which a shock can be transmitted by two main channels, trade and investment, represented by different layers of a multiplex network. From a practical point of view, a multiplex representation of global macroeconomic networks is obtained by coupling the WTW, forming the $T$ layer, with the CBIN, representing the $I$ layer. 

By considering the data sets available in Ref.~\cite{doi:10.1093/qje/qjt012}, GTI networks can be reconstructed for each year between 2001 and 2008. 
Figure~\ref{fig:hist_tot_w} shows the time evolution of the global value of traded goods $W_T$ and investment positions $W_I$, as defined by Eq. (1) of the main text.
One can see that both quantities  increase in time, but at different paces: while before the year 2000 total investment and trade were comparable, between 2001 and 2007 the total investment increased by more than three times, before decreasing in 2008 due to the financial crisis. The slower growth of global trade may be rooted in real-economy constraints, such as production capacity or shipping. 
The volume and exponential growth of global investment, on the contrary, demonstrates the need of including the financial layer in the study of international shock propagation.

GTI networks are directed, weighted, multiplex networks. 
Nodes represent countries, links in the $T$ layer represent exports/imports, 
links in the $I$ layer represent portfolio investments (that can be equity and/or debt) between countries, committed by both public and private actors. 
The weight $w_{ij}^{\ell}$ thus represents the weight from node $i$ to node $j$ in layer $\ell$.
 In Table~\ref{tab:prop}, we report some properties of the network obtained by data corresponding to the year 2005, which we use in the main paper.
Both layers are weakly but not strongly connected,  with less than half of links being bidirectional 
 and a relatively small reciprocity value $\rho$. 
 Both layers are quite dense, with a large average degree $\av{k}$, 
 and rather homogeneous degree distribution $P(k)$.

Despite the fact that degrees are quite homogeneously distributed, weights in both layers are very heterogeneous, as revealed by the broad tailed form of the weight distribution, shown in Fig. \ref{fig:w_distr}.
 The weight distributions of both layers, $P_{T}(w)$ and  $P_{I}(w)$, are compatible with power-law forms, $P_{\ell}(w) \sim w^{-\gamma_{\ell}}$, 
    with slightly different exponents $\gamma_{I} \simeq 1.36$ and $\gamma_{T} \simeq 1.5$.
   The out- and in-strength of node $i$, defined as $s_{i, \ell}^{out}=\sum_{j} w_{ij}^{\ell}$ and $s_{i, \ell}^{in}=\sum_{j} w_{ji}^{\ell}$, represent the total assets held and liabilities issued by a country $i$ for the investment layer, $\ell=I$, while they represent the total of exports and imports of the same country $i$ for $\ell=T$. Figure \ref{fig:w_distr} shows that the distributions $P_{\ell}(s)$ 
    are also heavy tailed, although noisy, due to the small size of the network. All distributions are compatible with  power-law forms, $P_{\ell}(s) \sim s^{-\gamma_{\ell}}$, with exponents 
    $\gamma_{\ell} \in [1.1, 1.3]$.

 \begin{figure}[tbp]
  \begin{center}  
    \includegraphics[width=8cm,angle=0]{\FigPath/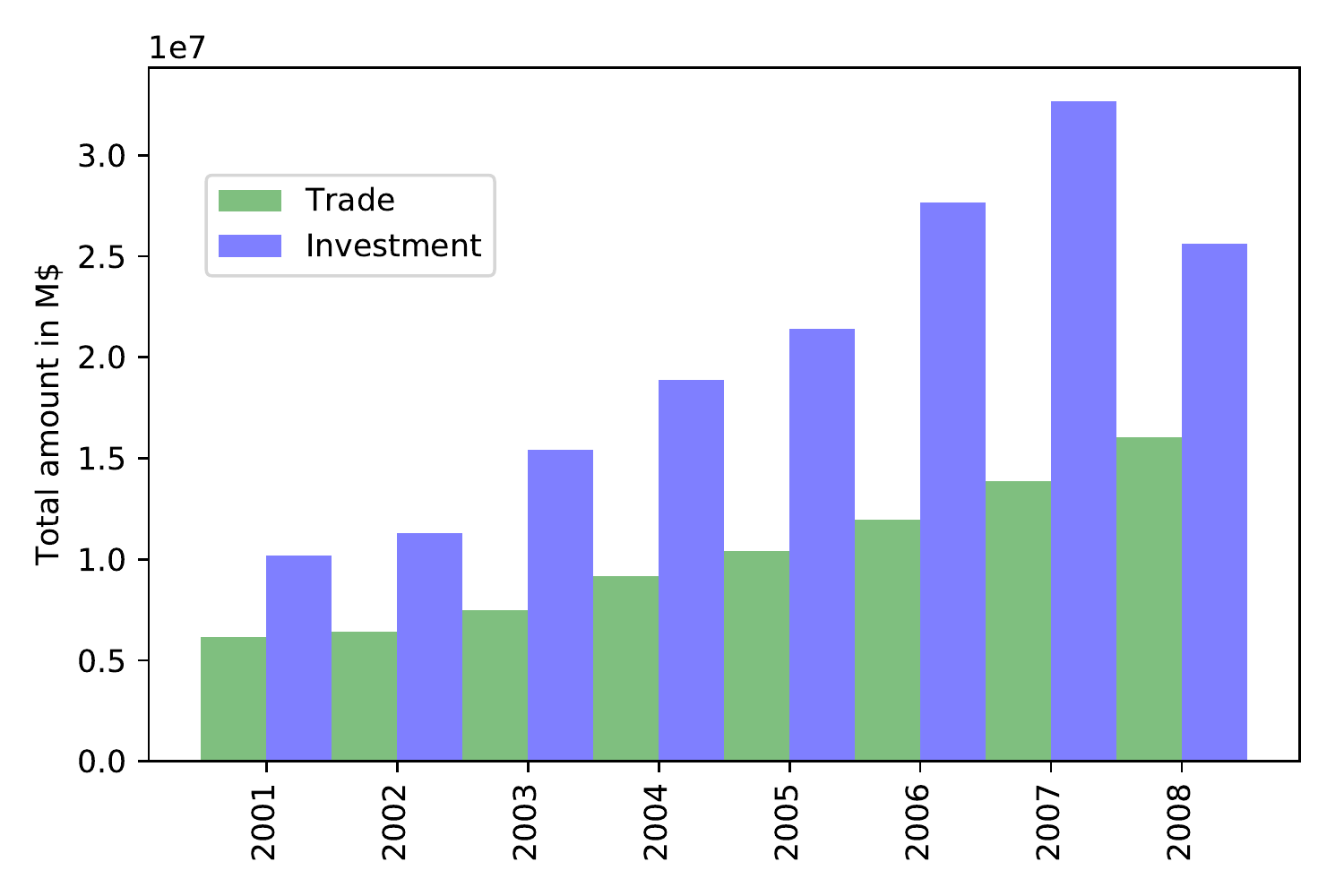}
      \end{center}
      \caption{ Global value of trade and investment from 2001 to 2008   \label{fig:hist_tot_w} 
      }
\end{figure}

\newpage

\begin{table}[tbp]
\centering
\caption{Some topological properties of GTI multiplex: number of nodes $N$, number of directed $E_{\ell}$ and overlapped $E_O$ edges, total weight $W_{\ell}$ (expressed in $10^{12}$ dollars), in trade $\ell=T$ and investment $\ell=I$ layers. }
\begin{tabular}{c|cc|cc|c}
$N$ & $E_T$ & $E_I$ & $E_O$ & $W_T$ & $W_I$ \\
\hline
186 & 12540 & 4499 & 3617  & 10.4 & 21.4 \\
\hline
\end{tabular}
\label{tab:prop}
\end{table}

\begin{figure}[tbp]
  \begin{center}  
    \includegraphics[width=8cm,angle=0]{\FigPath/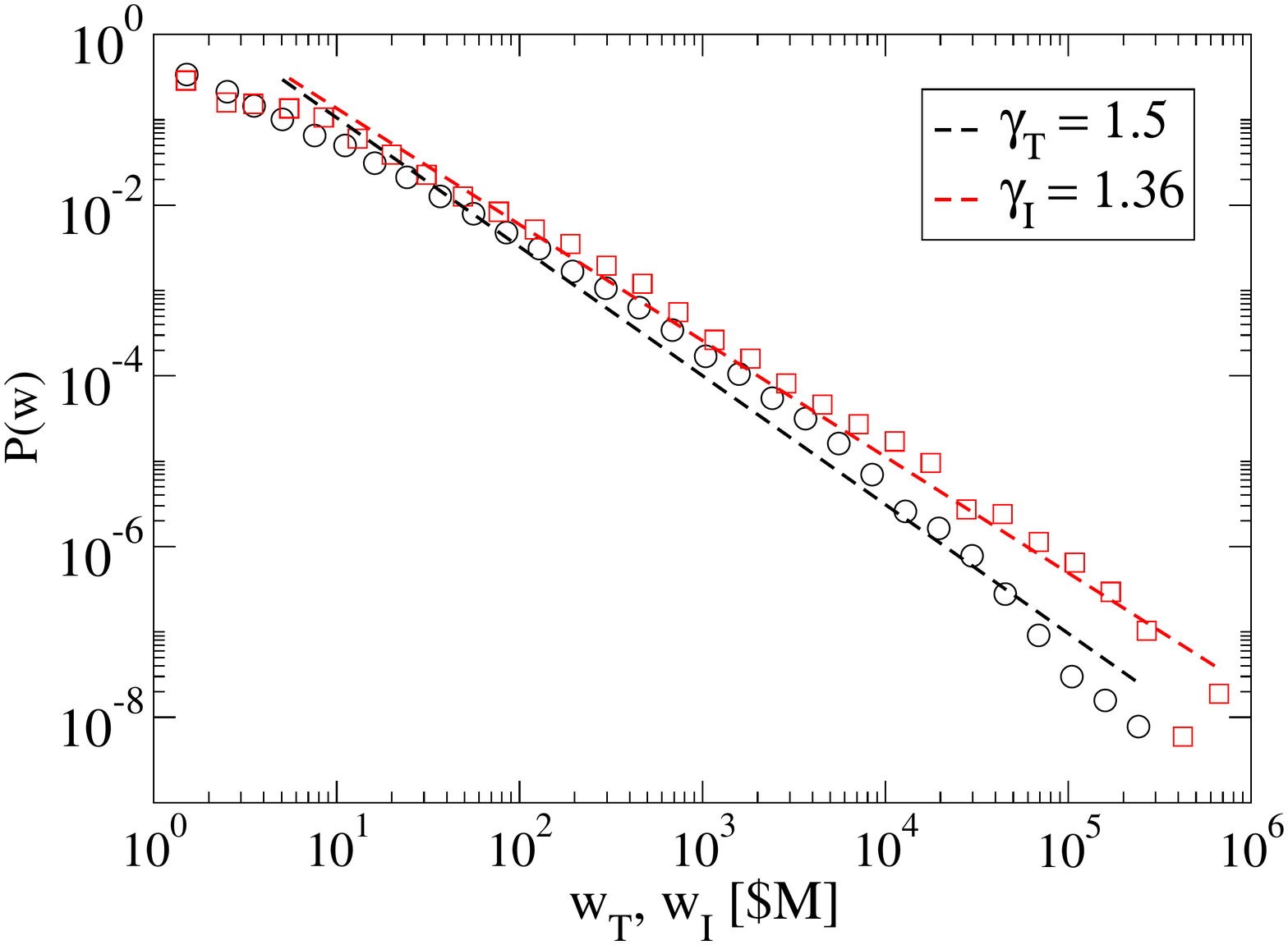}
    \includegraphics[width=8cm,angle=0]{\FigPath/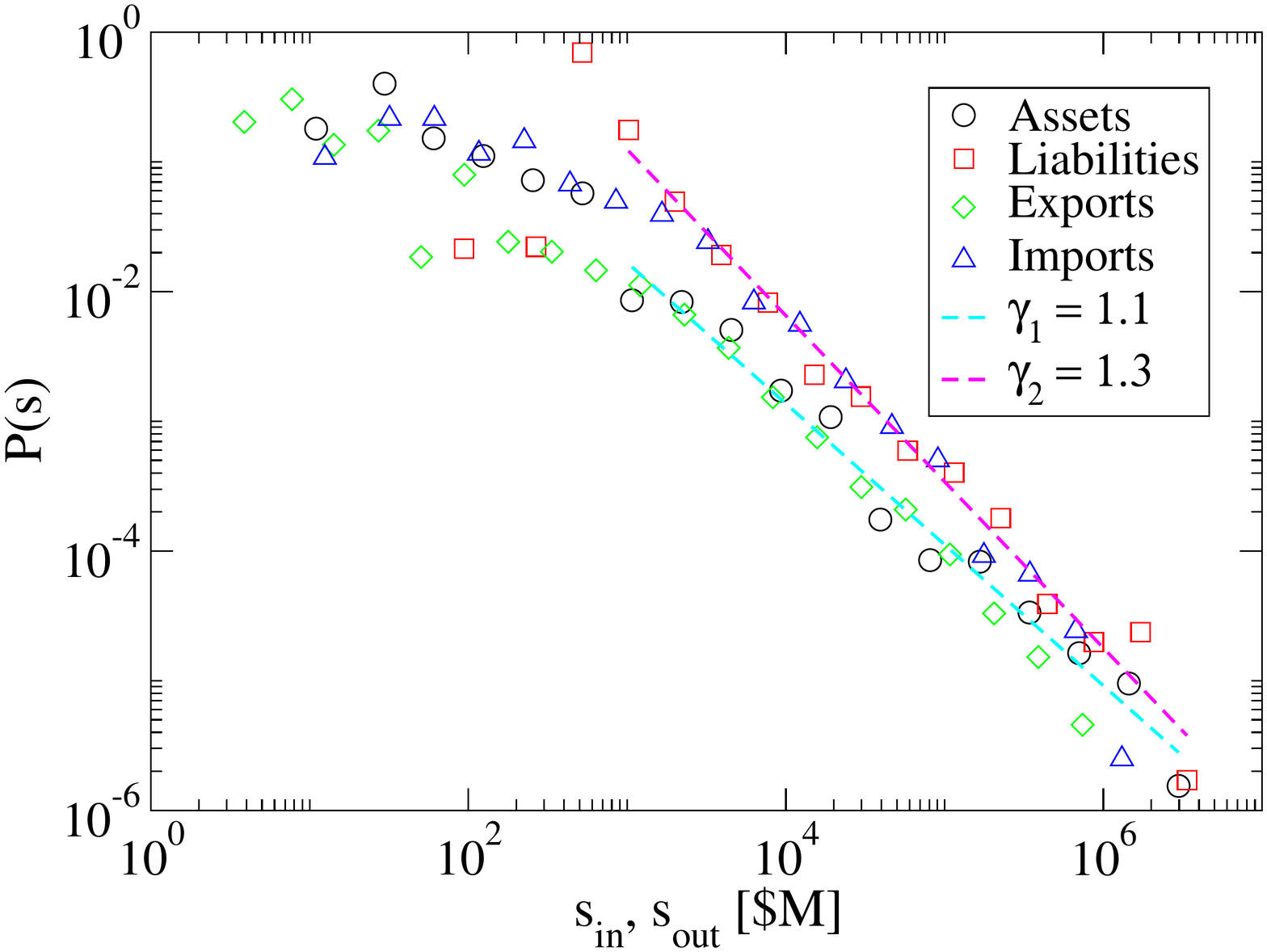}
      \end{center}
      \caption{ Probability distribution of weights $w_{ij}^{\ell}$ (left), and in- and out-strength  $s_{j}^{\ell}$ for layers $\ell = T$ and $\ell = I$ of the multiplex macroeconomic networks. 
      Power law functional forms $P(w) \sim w^{-\gamma}$ and $P(s) \sim s^{-\gamma}$ are drawn as a reference.  
  \label{fig:w_distr}}
\end{figure}


\newpage

\section{SIR dynamics of the shock propagation from distressed to neighboring countries}
\label{sec:detail_shock_prop}

In this section, we describe the details of the contagion dynamics of the shock propagation model, giving a concrete example to show how the economic distress spreads over the network.
First of all, let us fix the notation. Each link between two nodes $i$ and $j$ of the multiplex network is characterized by four macroeconomic quantities: goods exported from $i$ to $j$, $x_{ij}$, and goods imported from $i$ to $j$, $m_{ij} \equiv x_{ji}$, in the $T$ layer, assets held by $i$ and issued by $j$, $a_{ij}$, and liabilities issued by $i$ and held by $j$, $l_{ij} \equiv a_{ji}$, in the $I$ layer. In the same way, each node $i$ is characterized by four aggregated quantities: its total exports $X_i$, total imports $M_i$, total assets held $A_i$, total liabilities issued $L_i$. 
The dynamics of the network is represented by allowing these quantities to vary in time, adding explicitly the time-dependency: $x_{ij}(t)$,  $a_{ij}(t)$, $M_{i}(t)$, $L_{i}(t)$, etc. The notation $\delta Y_i(t) \equiv \frac{Y_i(t) - Y_i(t-1)} {Y_i(t-1)}$ stands for the relative variation of $Y_i(t)$ in time, $Y_i = \{X_i, M_i, A_i, L_i \}$,
where time $t$ is accounted by discrete time steps {in the shock propagation process,} $t=0,1, \ldots, t_{end}$. 
Equivalently, $\delta y_{ij}(t)$ represents the relative variation of quantity  $y_{ij} = \{x_{ij}, m_{ij}, a_{ij}, l_{ij} \}$  in time.

The contagion dynamics of the shock propagation model is simulated by means of a a Susceptible-Infected-Recovered (SIR) model, that allows to properly address reverberation and second order effects. It is important, indeed, to consider back and forth effects, since the epicenter country $i$, as well as any other node, can be hit back by the distress propagation. A node $i$, indeed, propagates the distress to a neighbor $j$ through links $x_{ji}(t)$ and $l_{ji}(t)$, and subsequently, node $j$ propagates its distress to all his neighbors, including node $i$ itself, though links $x_{ij}(t)$ and $l_{ij}(t)$. Moreover, each node that already propagated the distress will be further hit back by other neighboring nodes, as soon as the distress reaches them. 

In the SIR model, at each time $t$, each node $i$ can be in one of three states:
\begin{itemize}
\item  {\bf Susceptible, $s_i(t) = S$:} the country can receive distress from its neighbors, $\delta X_i(t) \neq 0$, $\delta L_i(t) \neq 0$, but it has not propagated it yet,  $\delta M_i(t)  = \delta A_i(t) = 0$;
\item {\bf Infected, $s_i(t) = I$:} the country propagates the distress accumulated to his neighbor by applying Eq. (2) of the main text, and it is characterized by $\delta M_i(t) \neq 0$ and $\delta A_i(t) \neq 0$;
\item {\bf Recovered or inactive, $s_i(t) = R$:} the country can receive distress from its neighbors, $\delta X_i(t) \neq 0$, $\delta L_i(t) \neq 0$, but it does not propagate it anymore.
\end{itemize}
The SIR dynamics can be summarized as follows. The following two steps are repeated in loop until no more infected node are present:
\begin{itemize}
\item {\bf Step 1}, each node $i$ in an infected state at time $t$, $s_i(t)=I$ (with $\delta M_i(t) \neq 0$ or $\delta A_i(t) \neq 0$) propagates the distress to all its neighbors (regardless of their status) and it becomes inactive immediately after, $s_i(t+1)=R$.
\item  {\bf Step 2}, each node $i$ in a susceptible state at time $t$, $s_i(t) = S$, with $\delta X_i(t) \neq 0$ or $\delta L_i(t) \neq 0$ (thus each node that received distress from any neighbor), applies Eq. (2) and becomes infected at time $t+1$, $s_i(t+1)=I$. After applying  Eq. (2), he sets $\delta X_i(t+1) = \delta L_i(t+1) = 0$. 
\end{itemize}
Step one and step two are repeated in loop until at some time $t^*$ each node $i$ will be in a susceptible (with $\delta X_i(t^*) = \delta L_i(t^*) = 0$) or inactive state (with $\delta X_i(t^*) \neq  0$  or $\delta L_i(t^*) \neq  0$).

It is worth to describe the distress propagation dynamics over the networks by means of a concrete example.
  Let us assume that at $t=0$, a shock originates in an epicenter country $i$. which, at time $t=1$, reduces its imports by a factor $\alpha_1$ and its investment in foreign assets by a factor $\beta_1$. For the sake of clarity in what follows, here we consider $|\alpha_1|>0$ and $|\beta_1|>0$ as the absolute value of the initial negative variation, while in the main text the initial variation are indicated by $\alpha$ and $\beta$ and can have any sign.
  
\begin{itemize}
\item  
  At $t=1$, node $i$ is in a infected state with $M_i(t=1) = (1-\alpha_1) M_i(t=0)$ and $A_i(t=1) = (1-\beta_1) A_i(t=0)$, while all other nodes $j$ are susceptible, $s_j(t=1)= S \quad \forall j$.  
  Since reductions in imports and assets are distributed proportionally among the neighboring nodes, the weight of each link from $j$ to $i$ is reduced by the same factor $\alpha_1$ and $\beta_1$, that is, $m_{ij}(t=1) = (1-\alpha_1) m_{ij}(t=0)$ and  $a_{ij}(t=1) = (1-\beta_1) a_{ij}(t=0)$. 
  Since imports from country $i$ to country $j$ are equal to exports from country $j$ to country $i$, each neighbor $j$ reduces its export to country $i$ as $x_{ji}(t=1) = (1-\alpha_1) x_{ji}(t=0)$, and the same applies for liabilities, $l_{ji}(t=1) = (1-\beta_1) l_{ji}(t=0)$. 
  This implies that each neighbor $j$ is forced to reduce its total exports by a different factor $\alpha_1^j$, $X_j(t=1) = (1-\alpha_1^j) X_j(t=0)$, and its total liabilities by a different factor $\beta_1^j$, $L_j(t=1) = (1-\beta_1^j) L_j(t=0)$, or equivalently $\delta X_1^j = - \beta_1^j$ and $\delta L_1^j = - \alpha_1^j$, in the notation of the main text. 
  Factors $\alpha_1^j \leq \alpha_1$ and $\beta_1^j \leq \beta_1$ depend on how important is country $i$ as economic partner for country $j$. 
  In the case limit of node $i$ being the only neighbor of node $j$ in both layers, it holds $\alpha_1^j = \alpha_1$ and $\beta_1^j = \beta_1$. 
  At this point, all infected nodes (in this case, only the epicenter country) have propagated the economic distress, thus step one of the loop is concluded. 
\item
In the next step, $t=2$, the epicenter country $i$ is set to a inactive state, $s_i(t=2) = R$, and  all neighboring nodes $j$ in a susceptible state that received the distress, with $\delta X_j(t=1) \neq 0$ or  $\delta L_j(t=1) \neq 0$, become infected, $s_j(t=2) = I$. 
 The variations of imports and asset investment of each country $j$, $\delta M_j(t=2)$ and $\delta A_j(t=2)$
are obtained through Eq. (2) of the main text, as a function of the variation of export and liabilities incurrence in the previous time step, $\delta M_j(t=2) = f(\delta X_j(t=1), \delta L_j(t=1))$ and $\delta A_j(t=2) = f(\delta X_j(t=1), \delta L_j(t=1))$. 
These variations depend on the set of propagation coefficients $c_{MX}$, $c_{ML}$, $c_{AX}$, and $c_{AL}$, and the noise terms. 
Immediately after applying Eq. (2) of the main text, each country $j$ sets $\delta X_j(t=2) =\delta  L_j(t=2) = 0$, to receive the next round of economic distress from his neighbors. 
   At this point, all susceptible nodes have applied Eq. (2), thus step two of the loop is concluded.
\item
We now repeat step one: each country $j$ in a infected state, $s_j(t=2) = I$ propagates the variations $\delta M_j(t=2)$ and $\delta A_j(t=2)$ proportionally to each neighbor $k$ through both trade and investment links, $\delta m_{jk}(t=2) = \delta M_j(t=2)$ and $\delta a_{jk}(t=2) =  \delta A_j(t=2)$. 
Note that the distress is propagated to all neighbors $k$, including the infected nodes (all the neighbors $j$ of the epicenter country) and the inactive nodes (only the epicenter country $i$). 
The total exports and liabilities of node $k$ are thus reduced by a different factors $\alpha_2^k$ and $\beta_2^k$, $ \delta X_k(t=2) = -\alpha_2^k$ and $\delta L_k(t=2) = -\beta_2^k$.  After all infected nodes (all neighbors $j$ of the epicenter country) have propagated the economic distress, step one of the loop is concluded. 
\item
Step two is repeated: At time $t=3$, all nodes $j$ previously infected become inactive, $s_j(t=3)= R$, and  nodes $k$ that received the distress, $\delta X_k(t=2) \neq 0$ or $\delta L_k(t=2) \neq 0$ that were susceptible, $s_k(t=2) = S$, become infected. 
That is, $s_k(t=3) = I$ only if $s_k(t=2) = S$ and ($\delta X_k(t=2) \neq 0 \lor  \delta L_k(t=2) \neq 0$). 
All nodes that become infected apply Eq. (2), obtaining variation of imports and asset investment $\delta M_k(t=3)$ and $\delta A_k(t=3)$ from variations $\delta X_k(t=2)$ and $\delta L_k(t=2)$ of the previous time step. 
Afterwords, each infected node $k$ sets $\delta X_k(t=3) = \delta L_k(t=3) = 0$, to receive the next round of economic distress from its neighbors, and step two is concluded again. 
\end{itemize}

Step one and step two are repeated in loop until at some point $t=t^*$, each node $i$ will be in a susceptible (with $\delta X_i(t^*) = \delta L_i(t^*) = 0$) or inactive state (with $\delta X_i(t^*) \neq 0$  or $\delta L_i(t^*) \neq  0$). At this point, the SIR dynamics is concluded, and the economic distress has propagated through all the network. Note that each node $i$ propagates the distress accumulated by applying Eq. (2) of the main text at most one time. 

It is important to note that at $t=t^*$, each inactive node $i$ is characterized by  $\delta X_i(t^*) \neq 0$  or $\delta L_i(t^*) \neq  0$, meaning that it is affected by second order effects of the contagion, that should be taken into account. For this reason, we repeat the whole SIR dynamics, by using the distress accumulated by inactive nodes as initial conditions for the new dynamics. Each inactive node $i$ at time $t^*$, $s_i(t^*)=R$, becomes infected at time $t^*+1$, $s_i(t^*+1) = I$, thus it applies Eq. (2), obtaining new variations in imports and asset investment, $\delta M_i(t^*+1)$ and $\delta A_i(t^*+1)$. 
By repeating the SIR dynamics, a second contagion wave spreads over the network, until again at some time $t=t^{**}$ each node will be in a susceptible or inactive state. 
We repeat the SIR dynamics a number $n$ of times, that ensures that the system has reached a steady state. 
The fact that at each reverberation the distress propagated is smaller ensures that the dynamics converges quickly.

It is important to note that the duration of the whole contagion dynamics, given by the $n$ repetitions of the SIR dynamics, is  a free parameter of the model. However, after a certain number of reverberations the system is stable and does not evolve in time anymore. Figure~\ref{fig:compT} shows the evolution in time of exports, imports, assets and liabilities, $X_i(t)$, $M_i(t)$, $A_i(t)$, and $L_i(t)$ for two countries, $i = $ UK and $i = $ Germany, with respect to the same quantities at $t=0$, the time at which the shock takes place in a single epicenter country, namely the United States, with initial conditions $\alpha = -0.1$, $\beta = -0.3$. One can see that after $n \simeq 20$ repetition steps the system does not evolve in time anymore. Note also that there are slight differences in the evolution by choosing a different number of repetitions $n$ (represented by continuous and dashed lines in Fig. \ref{fig:compT}). In the paper we set $n=50$ that ensures that the system has reached a steady state at time $T$ after the shock.

 \begin{figure}[tbp]
  \begin{center}  
    \includegraphics[width=8cm,angle=0]{\FigPath/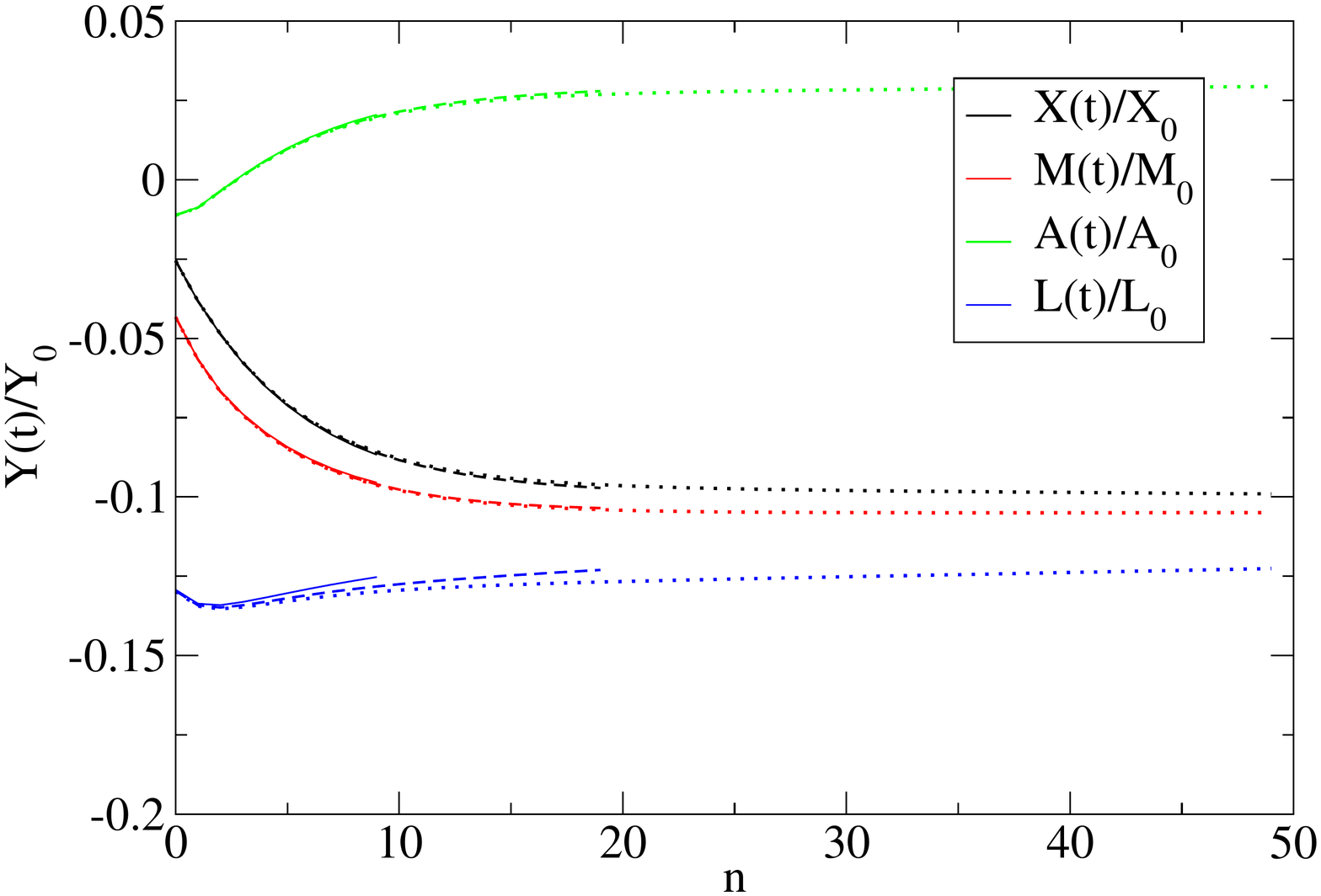}
    \includegraphics[width=8cm,angle=0]{\FigPath/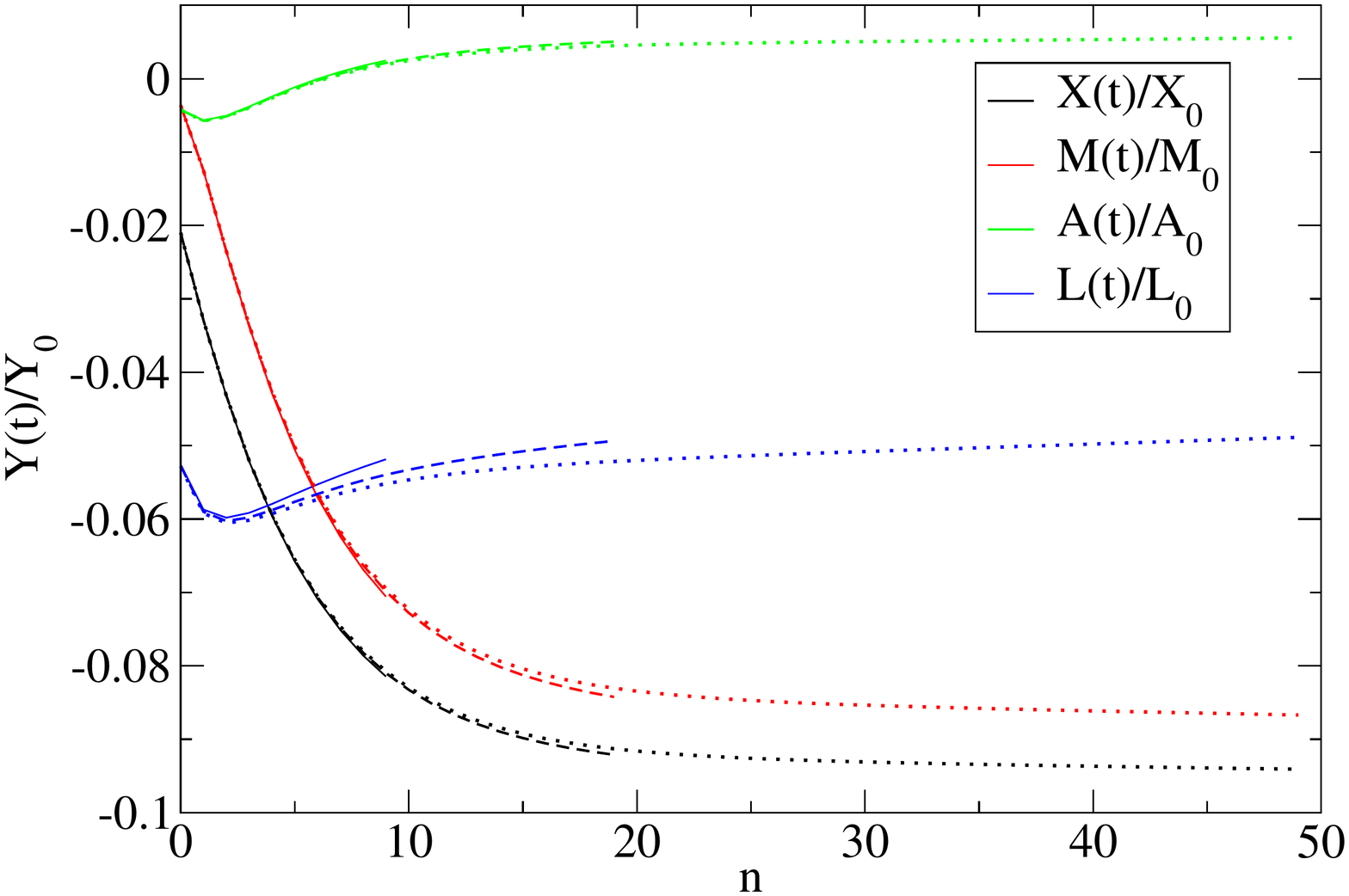}
      \end{center}
      \caption{ Evolution of exports, imports, assets and liabilities of the UK (left) and Germany (rigth) by using a different number of repetions $n$ of the SIR contagion dynamics. Results averaged over 100 runs.  \label{fig:compT} 
      }
\end{figure}

\newpage

\section{Estimation of the pass-through coefficients of shock propagation model }

In this section, we describe how we estimate pass-through coefficients of the shock propagation model, and we briefly discuss them. 
For each country, the trend terms, pass- through coefficients, and noise terms in Eq. (2) of the main text 
are estimated by calculating variances and co-variances of the four time series $\{dX_t, dM_t, dA_t, dL_t\}$. We consider yearly data from 1980 to 2015, by excluding recession periods (i.e. years 1982, 1991, 2008), as reported by the IMF, see Section \ref{sec:description-data}. 
The variance of the noise terms, $\sigma_{\epsilon_1}^2$ and $\sigma_{\epsilon_2}^2$, incorporates the reliability of the propagation coefficients and trend terms. If data for a country are scarce, with respect to one or several time series, and no clear relation emerges from two macroeconomic variables, then the variance of the corresponding noise term will be large if compared to the coefficient multiplied by the variance of the corresponding variable, e.g. $\sigma_{\epsilon_1}^2 \gg  c_1 \langle dL_t^2 \rangle$. 
In this case, such country would propagate only noise to the system. For this reason, if for any country, at time $t$, it holds e.g. $\sigma_{\epsilon_1}^2 \ge  b_1 \langle dX_t^2 \rangle$ or $\sigma_{\epsilon_1}^2 \ge  c_1 \langle dL_t^2 \rangle$, then we set $dM_t = 0$.
The same conditions apply for the term $dA_t$. 
This might underestimate the spillover effect, but it ensures the system to be stable. 
We check that the countries that do not fulfill conditions for stability are few. 

\begin{figure*}[tb]
  \begin{center}  
    \includegraphics[width=8.4cm,angle=0]{\FigPath/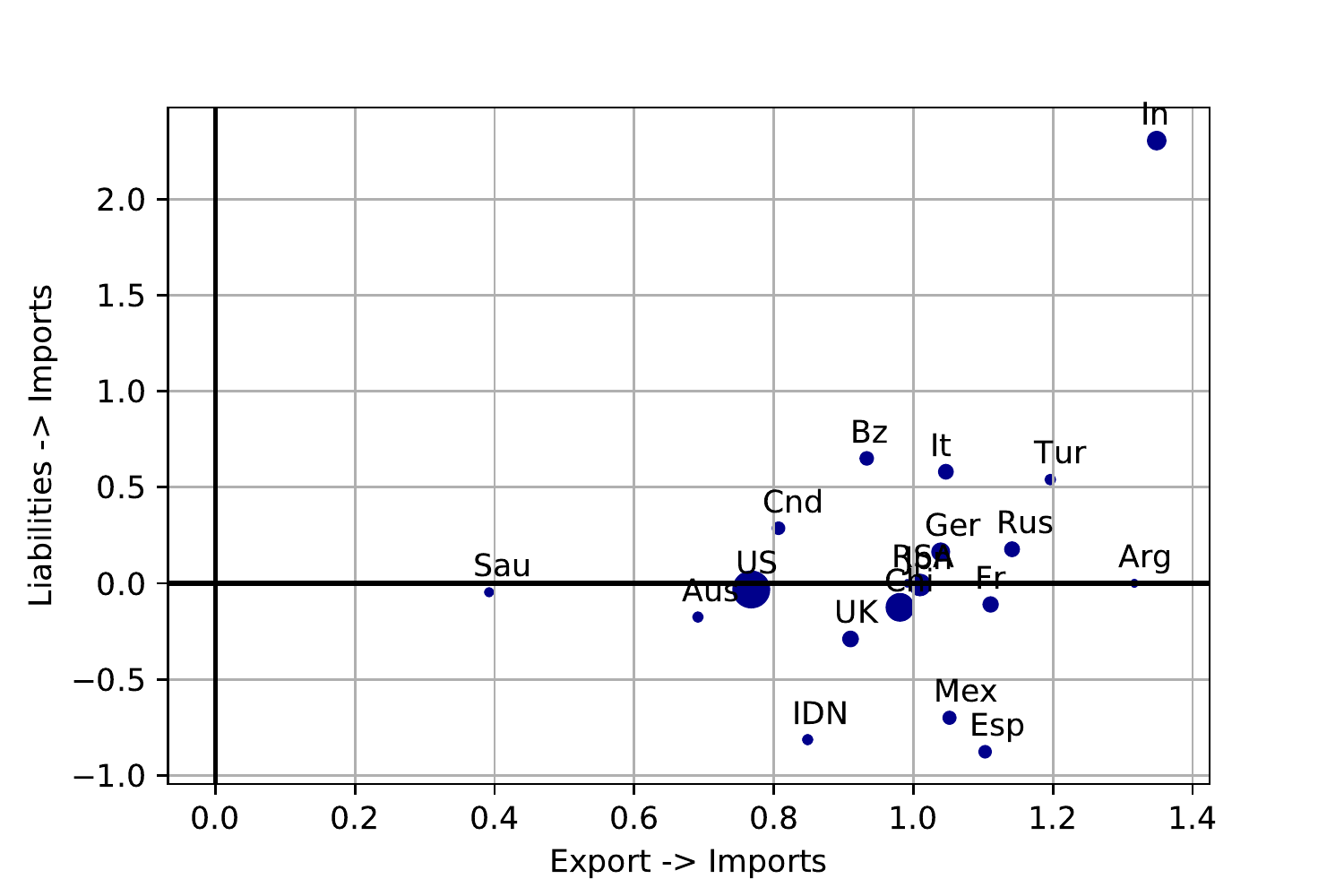}
    \includegraphics[width=8.4cm,angle=0]{\FigPath/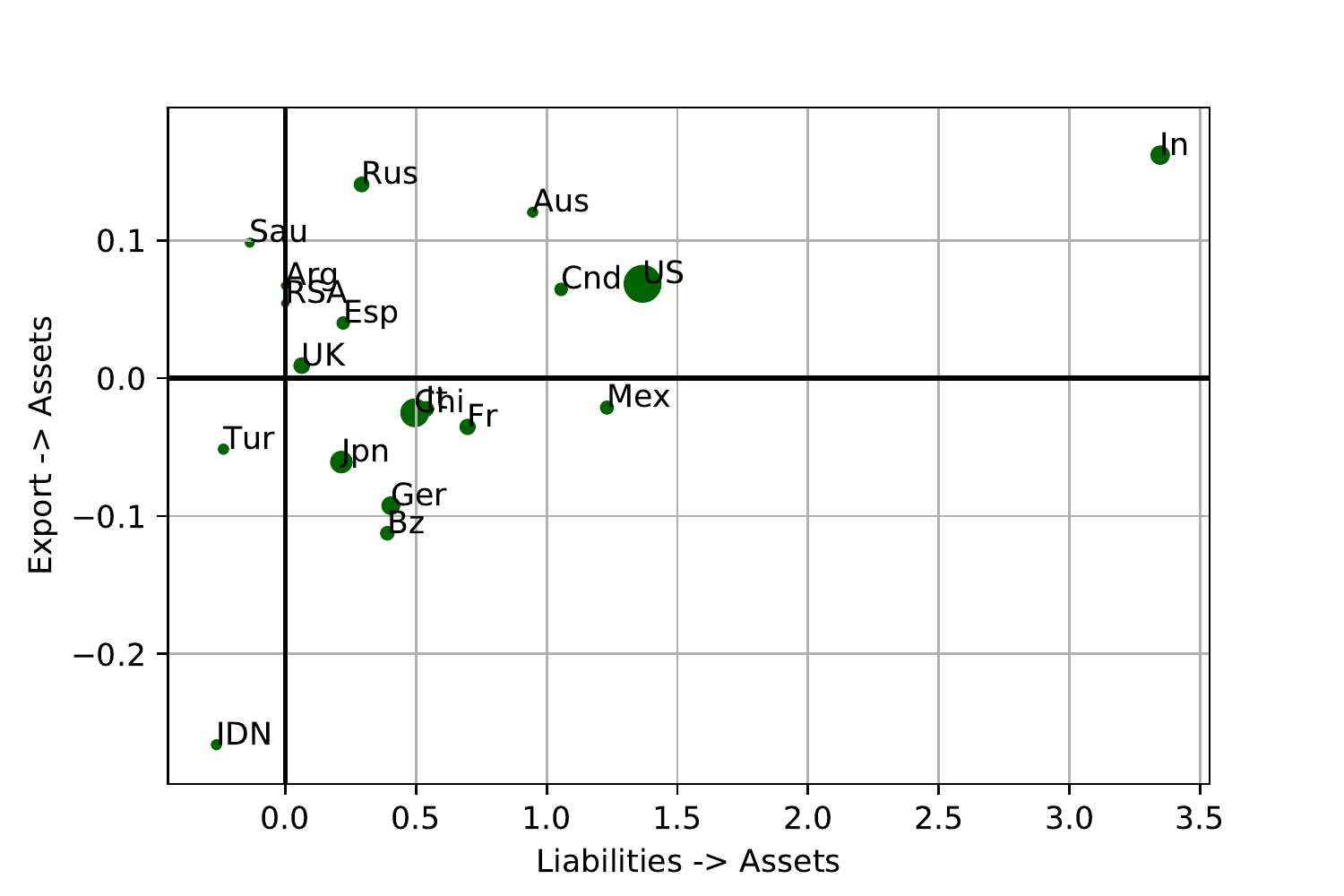}
      \end{center}
      \caption{ Scatter plots of intra-layer (x-axis) versus  inter-layer (y-axis) coefficients. Pass-through coefficients to trade ($c_{MX}$ on the x-axis and $c_{ML}$ on the y-axis), and to investment  ($c_{AL}$ on the x-axis and $c_{AX}$ on the y-axis),   for the $G_{20}$ group. The size of the countries is proportional to their GDP. 
 \label{fig:hist_prop_coeff}}
\end{figure*}

Figure \ref{fig:hist_prop_coeff} shows pass-through coefficients for countries belonging to the $G_{20}$ group.
The plot on the left shows propagation coefficients  $c_{MX}$ (export to import) and $c_{ML}$ (liabilities to import), while the plot on the right shows  propagation coefficients  $c_{AL}$  (liabilities to assets) and $c_{AX}$ (exports to assets).
The major contribution for the propagation of a shock on the GTI multiplex comes from \textit{intra-layer} coefficients acting within the same layer. For many countries, the {inter-layer} propagation $I \rightarrow T$, represented by coefficient $c_{ML}$, is  larger than the propagation $T \rightarrow I$, represented by coefficient $c_{AX}$.
For trade, Fig.\ref{fig:hist_prop_coeff} (left), the intra-layer term $c_{MX}$, representing the dependence  between imports and exports, is dominant with respect to the inter-layer term $c_{ML}$. If we exclude the case of India, it holds $|c_{MX}| > |c_{ML}|$ for all large economies, as expected. 
The coefficient $c_{MX}$ is bounded between $c_{MX} \in [0, 1.5]$, with most countries having $c_{MX} \simeq 1$, indicating a positive, strong correlation between imports and exports, as previous empirical findings showed. 
The most notable exception is Saudi Arabia with $c_{MX} \simeq 0.4$, a rich Middle East oil producer, whose imports are known to depend little on revenues from exports and financial assets.
The inter-layer term $c_{ML}$ is generally bounded between  $c_{ML} \in [-0.75, 0.75]$, indicating that the dependence between a variation in exports and the incurrence in liabilities can be positive or negative, depending on the country. 
For investment, Fig.\ref{fig:hist_prop_coeff} (right), the intra-layer term $c_{AL}$ is also dominant with respect to the inter-layer term $c_{AX}$. The coefficient between variation of assets and liabilities is generally positive, also bounded between $c_{AL} \in [0, 1.5]$ (excluding India). The inter-layer coefficient $c_{AX}$ is much smaller, bounded between $c_{ML} \in [-0.3, 0.2]$, indicating that the correlation between a variation in exports and assets acquisition is very weak. 
The dominance of intra-layer propagation terms, generally positive, is confirmed by simple considerations of balance of payments flows: a negative (positive) variation in export (liabilities) revenues is expected to generate a negative (positive) variation in imports (asset acquisition).

\newpage

\section{Vulnerability of countries to propagating shocks}

 The impact of an initial shock characterized by parameters $(\alpha,\beta)$, in an epicenter country $E$, to another country $i$,  can be quantified by considering the relative variations $\Delta Y_i({\alpha, \beta, E})$  of each macroeconomic quantity $Y_i = \{ X_i, M_i, A_i, L_i \}$ (exports, imports, assets acquisition and liabilities incurrence) of country $i$, as described by Eq. (3) of the main text.
Figure~\ref{fig:fit_distr_loss} shows the distribution of relative variations $\Delta Y_i({\alpha, \beta, E})$  for $i=$ Germany, obtained by a shock with parameters  $\alpha=-0.1$ and $\beta=-0.3$ and epicenter in the United States. 
 We find the best fit of these distribution with some standard fitting function, i.e. gamma, beta, Rayleigh, and normal functions, and extract the average, $\langle \Delta Y_i({\alpha, \beta, E}) \rangle$, and the Value-at-Risk, $VaR[ \Delta Y_i({\alpha, \beta, E}) ]$, for $p=0.05$ of each quantity, defined such that the probability of a negative variation greater than $VaR$ is less than or equal to $p$ while the probability of a loss less than $VaR$ is less than or equal to $1- p$. 

\begin{figure}[tbp]
  \begin{center}  
    \includegraphics[width=8.5cm,angle=0]{\FigPath/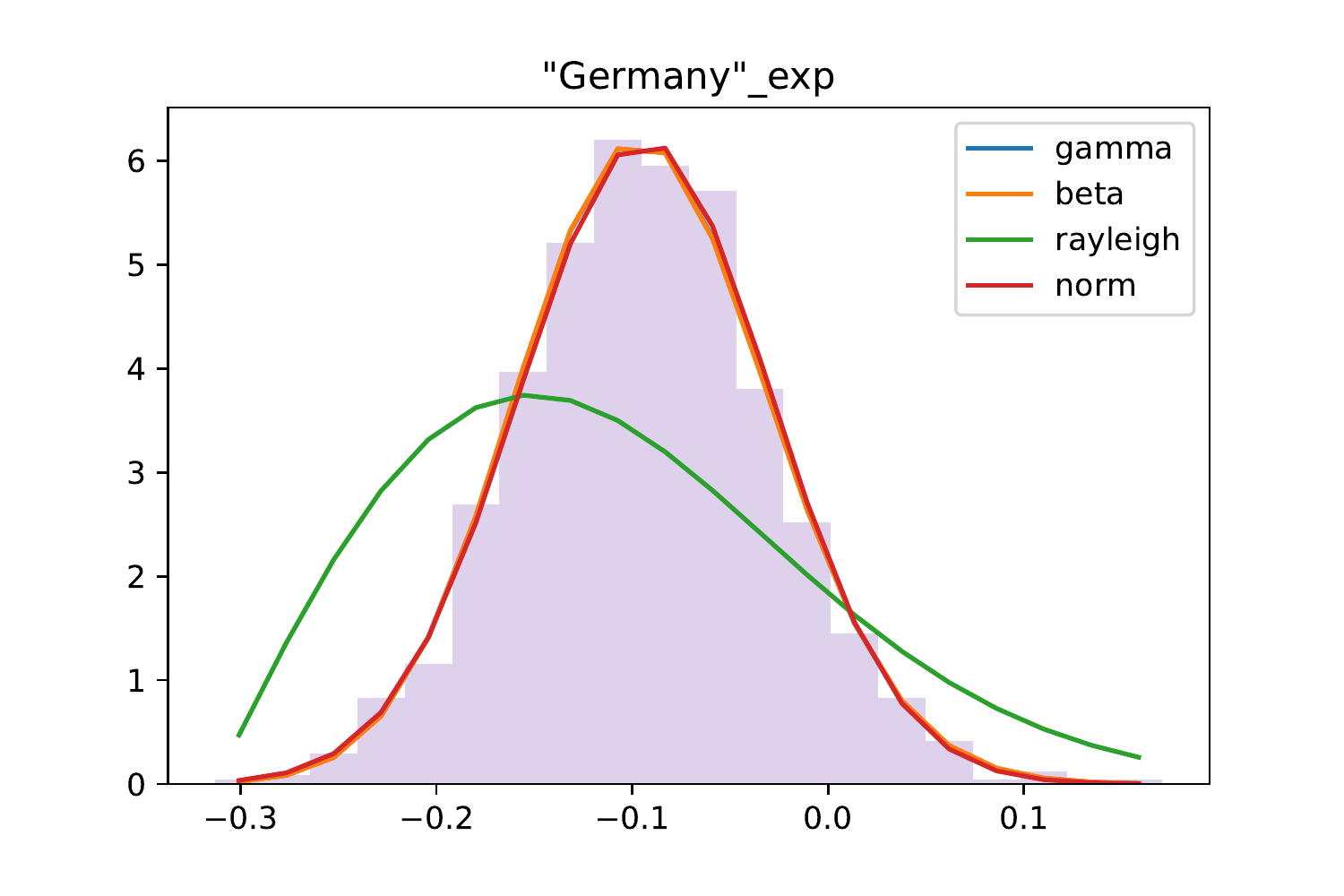}
    \includegraphics[width=8.5cm,angle=0]{\FigPath/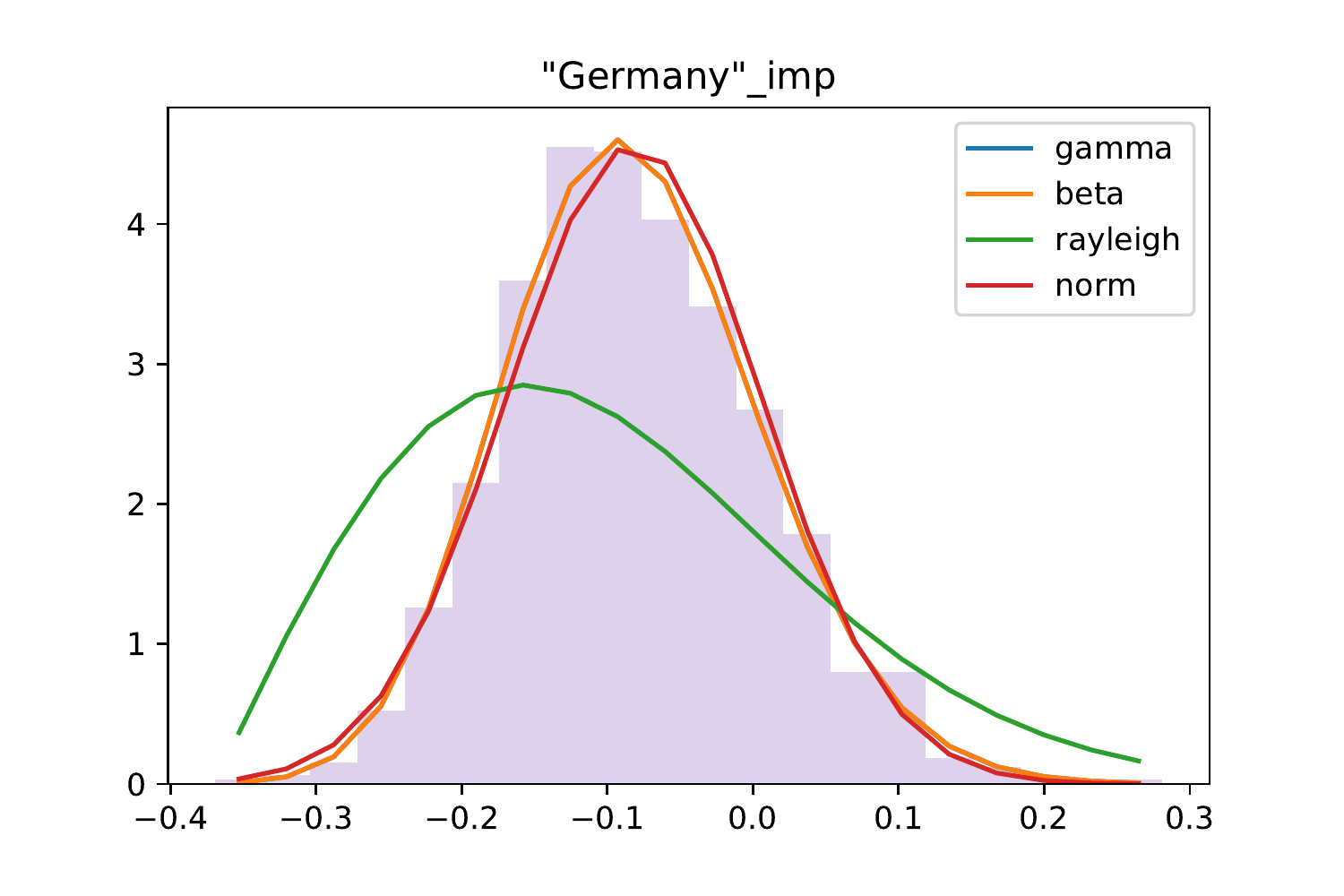}
    \includegraphics[width=8.5cm,angle=0]{\FigPath/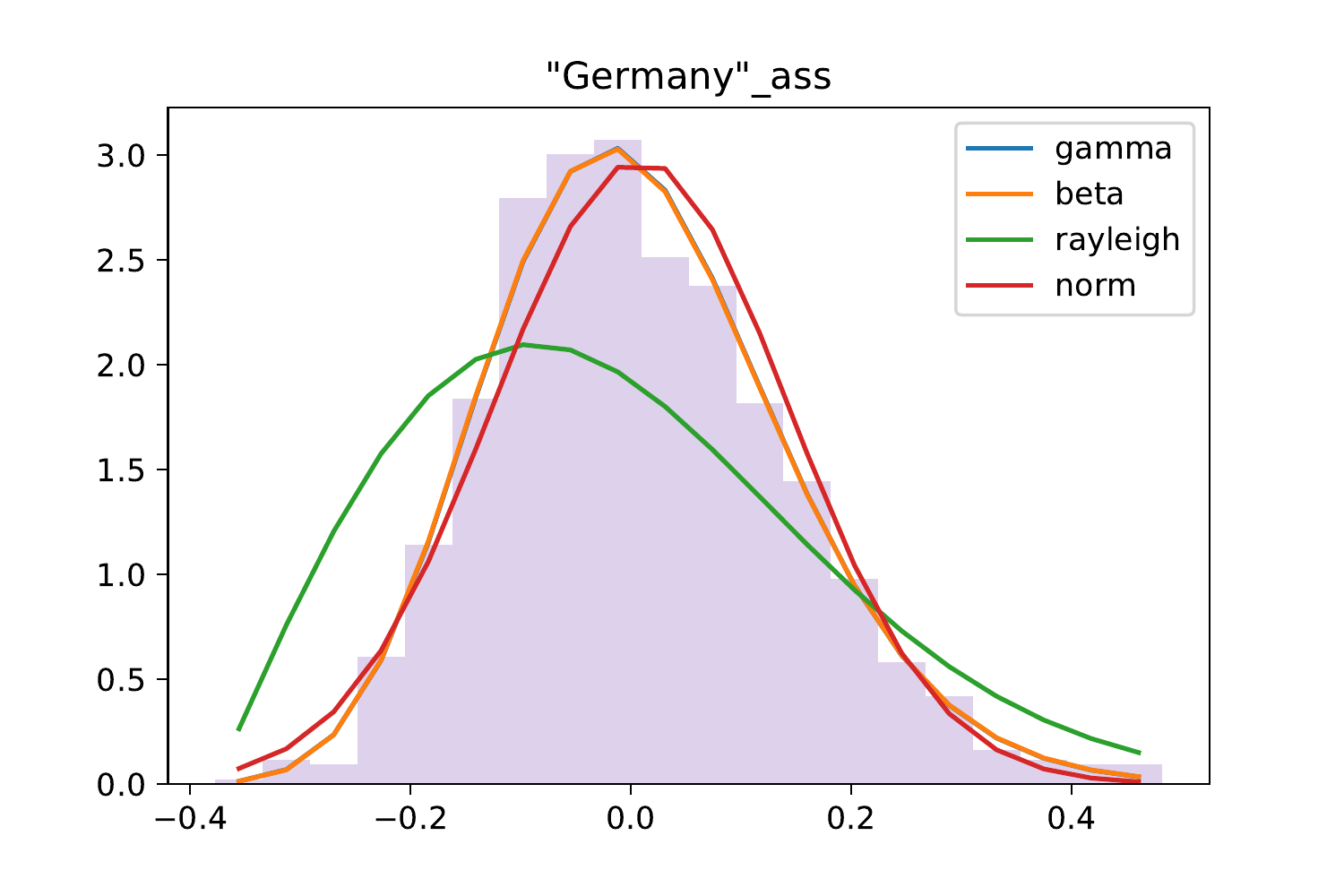}
    \includegraphics[width=8.5cm,angle=0]{\FigPath/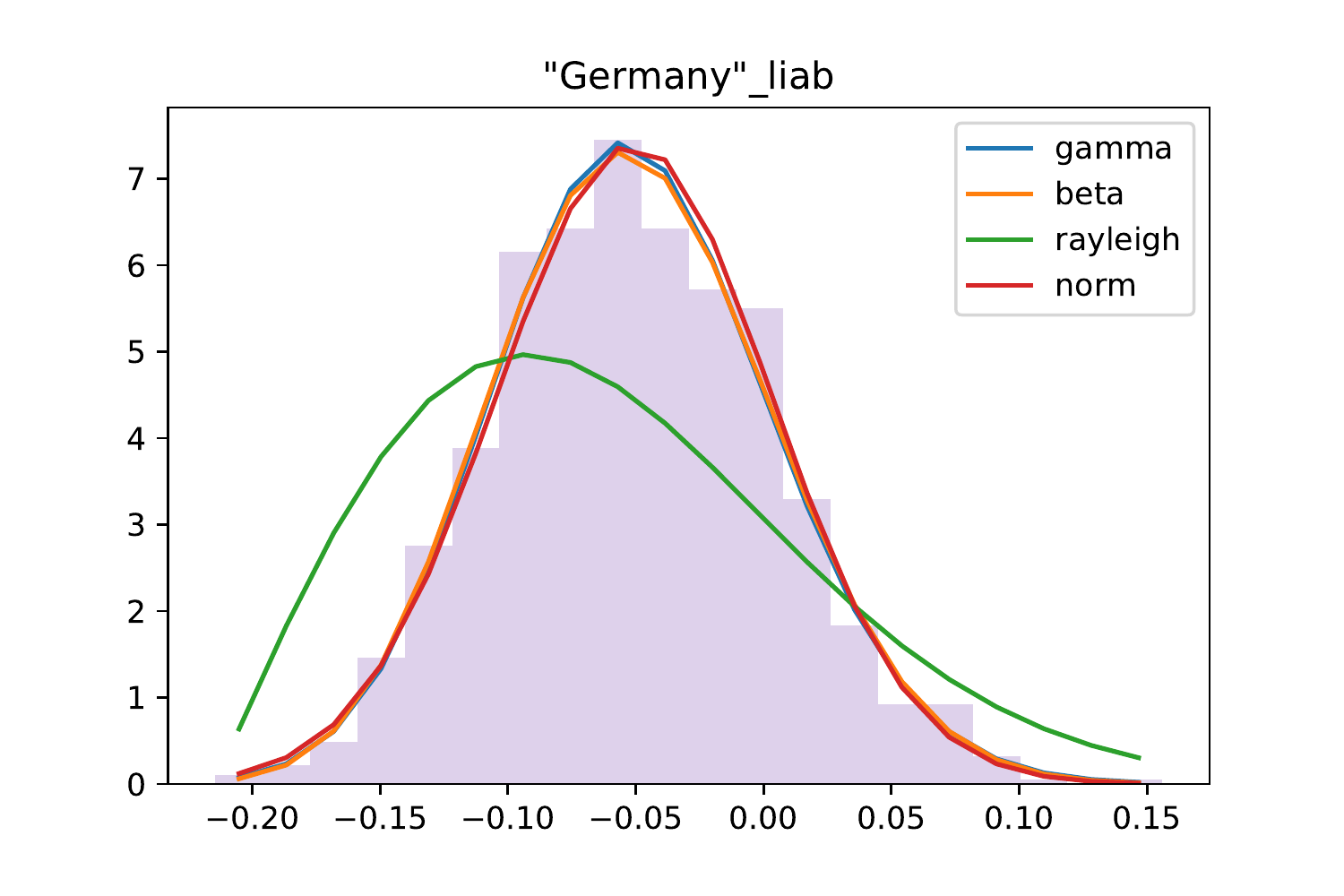}
      \end{center}
      \caption{ Distributions of the relative variations $\Delta X_i({\alpha, \beta})$ (top, right), $\Delta M_i({\alpha, \beta})$ (top, left), $\Delta A_i({\alpha, \beta})$  (bottom, left), and $\Delta L_i({\alpha, \beta})$ (bottom, right), for parameters $\alpha=-0.1$ and $\beta=-0.3$. Fitting functions gamma, beta, Rayleigh and normal functions are plotted as continuous lines with different colors. \label{fig:fit_distr_loss}}
\end{figure}

Figure \ref{fig:maps} shows the average vulnerability of exports, $\av{ V_i(X_i)}$ (left plots), and  incurrence in liabilities, $\av{ V_i(L_i)}$, of each country $i$ with respect to a shock originated in the United States (first row), or in China (second row), or in countries belonging to the EZ (third row), characterized by  $\alpha=-0.4$ and $\beta=-0.1$. There maps correspond to Fig.~1 of the main text, which shows the $VaR$ of the same quantities, for the same initial shock. One can see that the average vulnerability shows the same qualitative behavior across different countries of the corresponding $VaR$.

\begin{figure*}[tbp]
  \begin{center}  
    \includegraphics[width=17cm,angle=0]{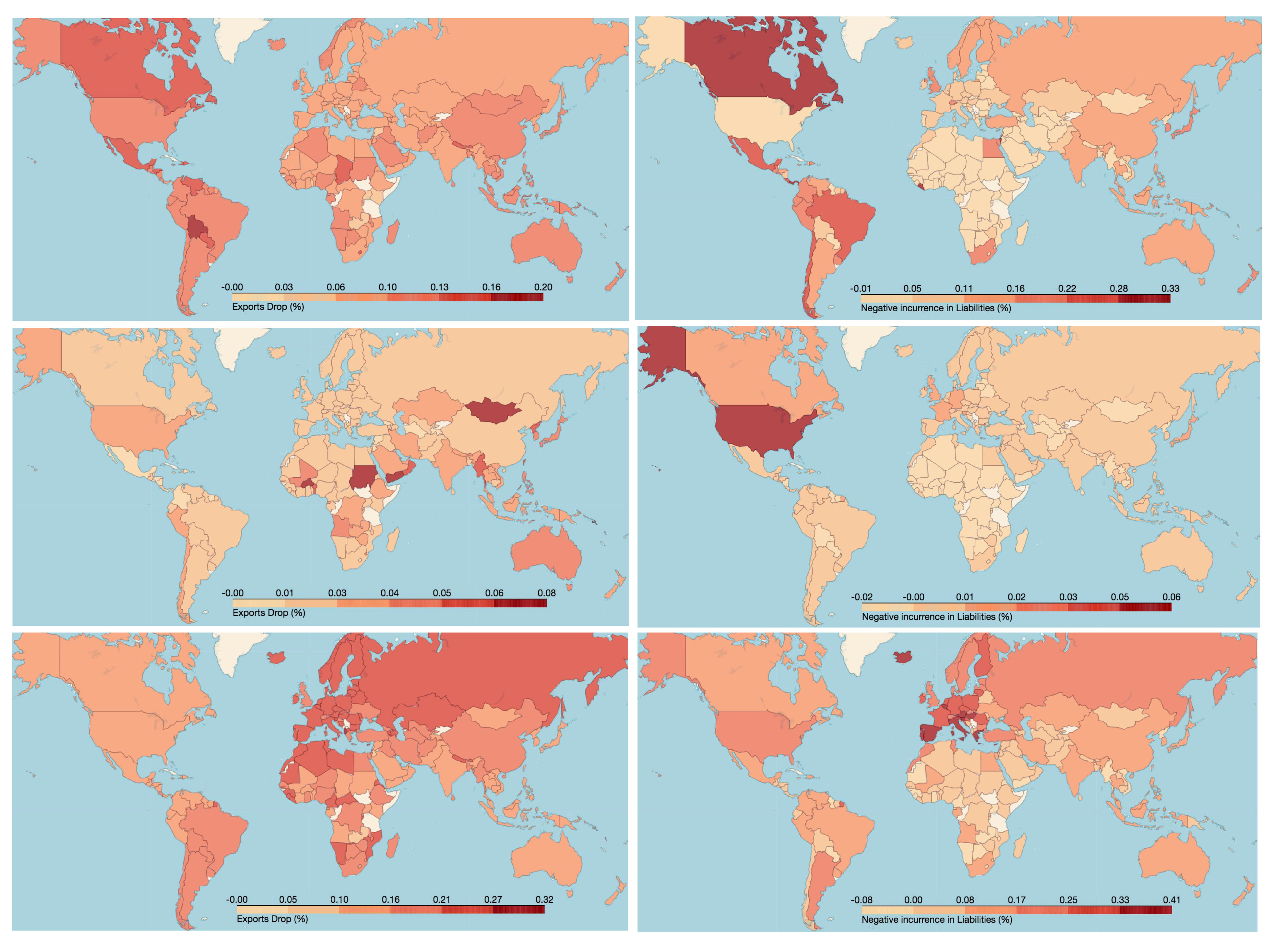}
      \end{center}
      \caption{ 
   Vulnerability of each country with respect to a shock originated in the United States (first row), China (second row), or in countries belonging to the EZ (third row), characterized by  $\alpha=-0.4$ and $\beta=-0.1$.   
      Colors indicate the average of exports, $VaR[\Delta X_i ]$ (left plots), and of incurrence in liabilities, $VaR[\Delta L_i]$ (right plots).  
         \label{fig:maps}}      
\end{figure*}

\newpage

\section{Quantifying systemic impact of epicenter countries}

Here we show the systemic impact as a function of the magnitude of the initial shock (Fig. \ref{fig:sys_imp}), and the deviations from the linear relation (Fig. \ref{fig:residuals}). These Figures correspond to Fig.~3 of the main text, with different values of $\alpha$ and $\beta$.

\begin{figure}[tbp]
  \begin{center}  
    \includegraphics[width=8.5cm,angle=0]{\FigPath/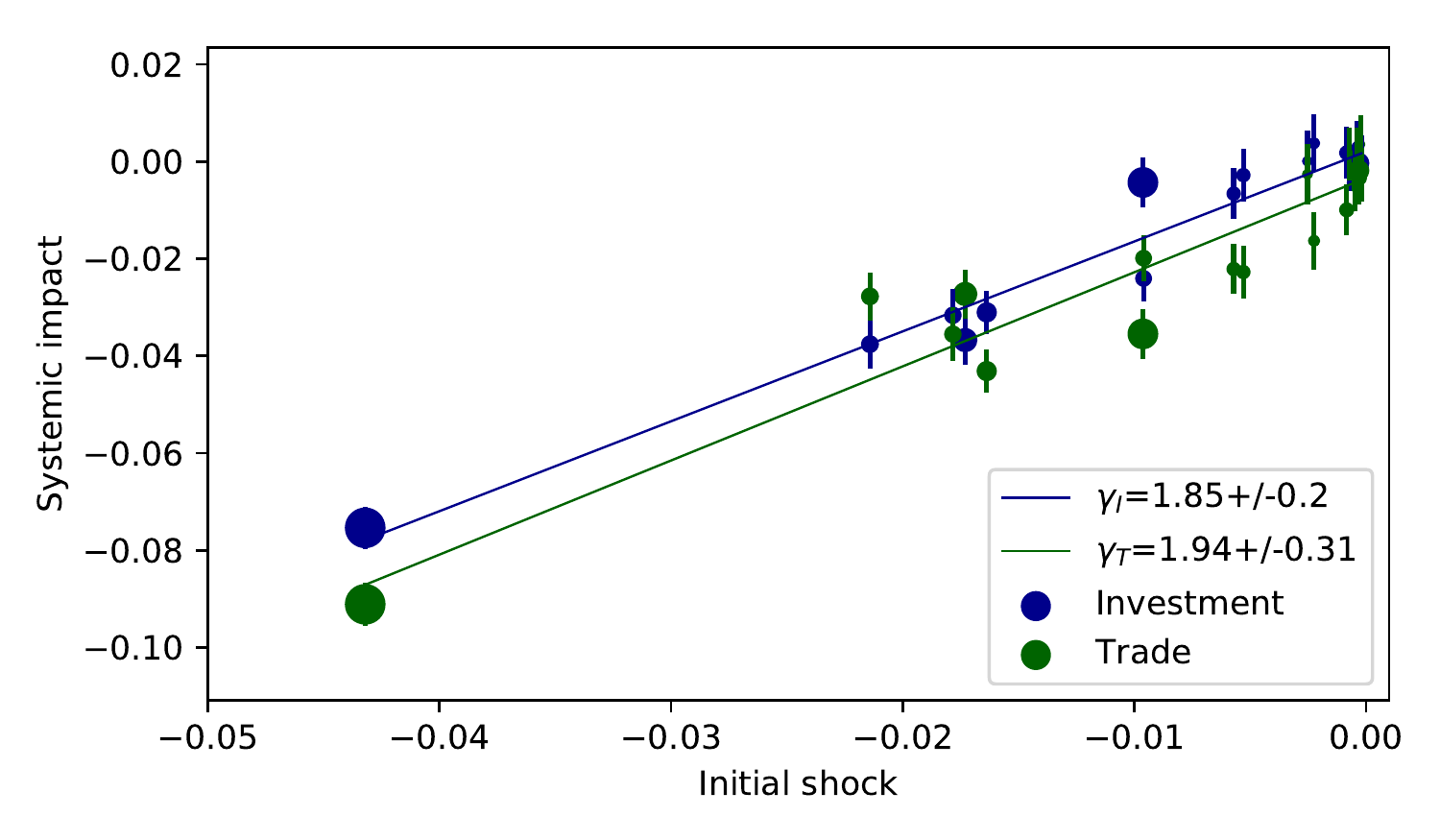}    
    \includegraphics[width=8.5cm,angle=0]{\FigPath/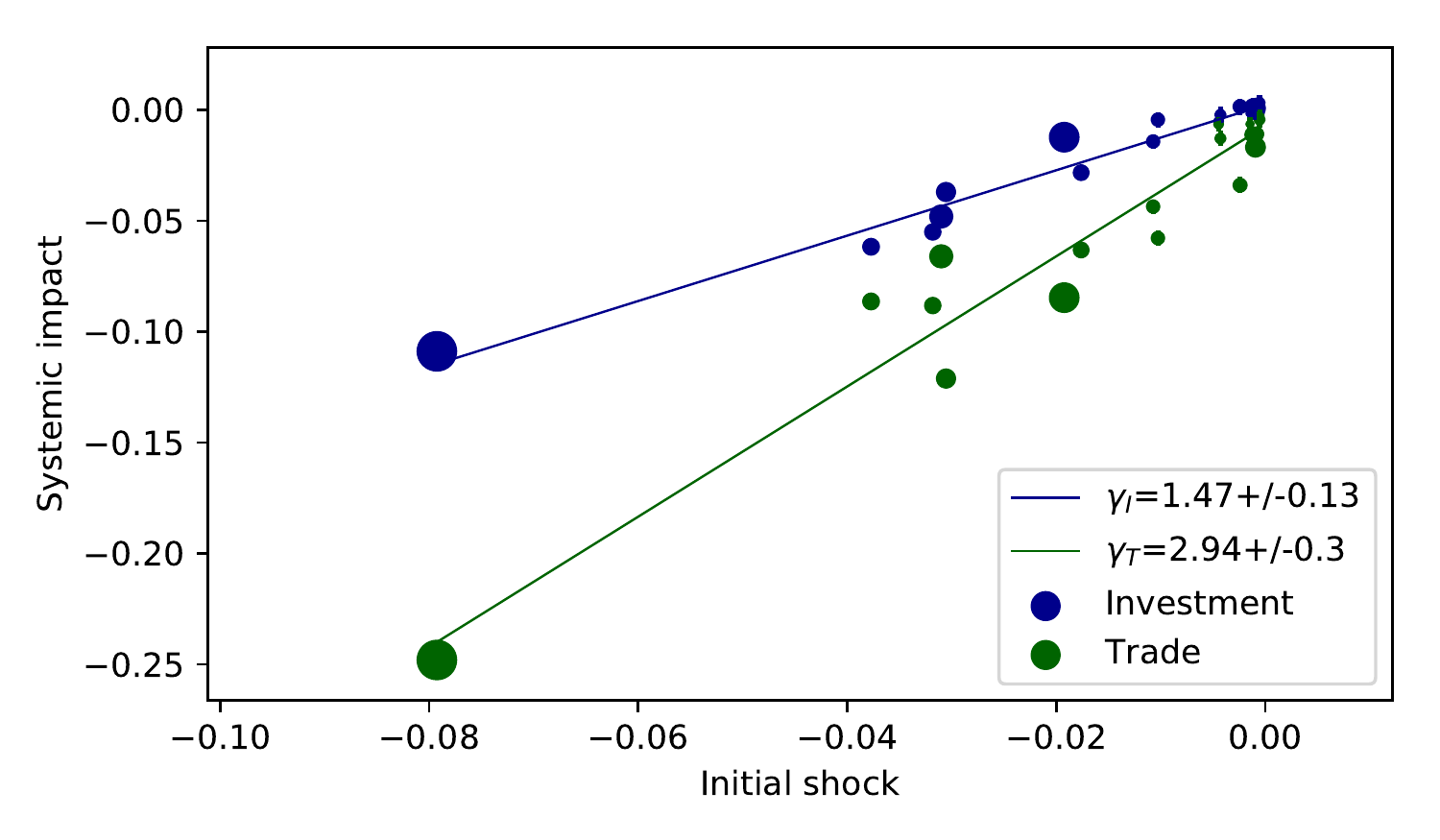}  
      \end{center}
      \caption{ 
   Systemic impact on global trade $\mathcal{S}_i^T$ and investment $\mathcal{S}_i^I$, as a function of the magnitude of the initial shock $\mathcal{I}_i/ (W_I + W_T)$. 
   The initial shock is characterized by $\alpha = -0.1, \beta = -0.3$ (left), or $\alpha =-0.3, \beta = -0.5$ (right). Countries belonging to the $G_{20}$ group are shown. Error bars represent the standard error of the mean for  $\mathcal{S}_i$. Regression coefficients $\gamma_{\ell}$ are plotted with $95\%$ CI. Size of dots is proportional to countries' GDP.   \label{fig:sys_imp} }
\end{figure}

\begin{figure}[tbp]
  \begin{center}  
    \includegraphics[width=8.5cm,angle=0]{\FigPath/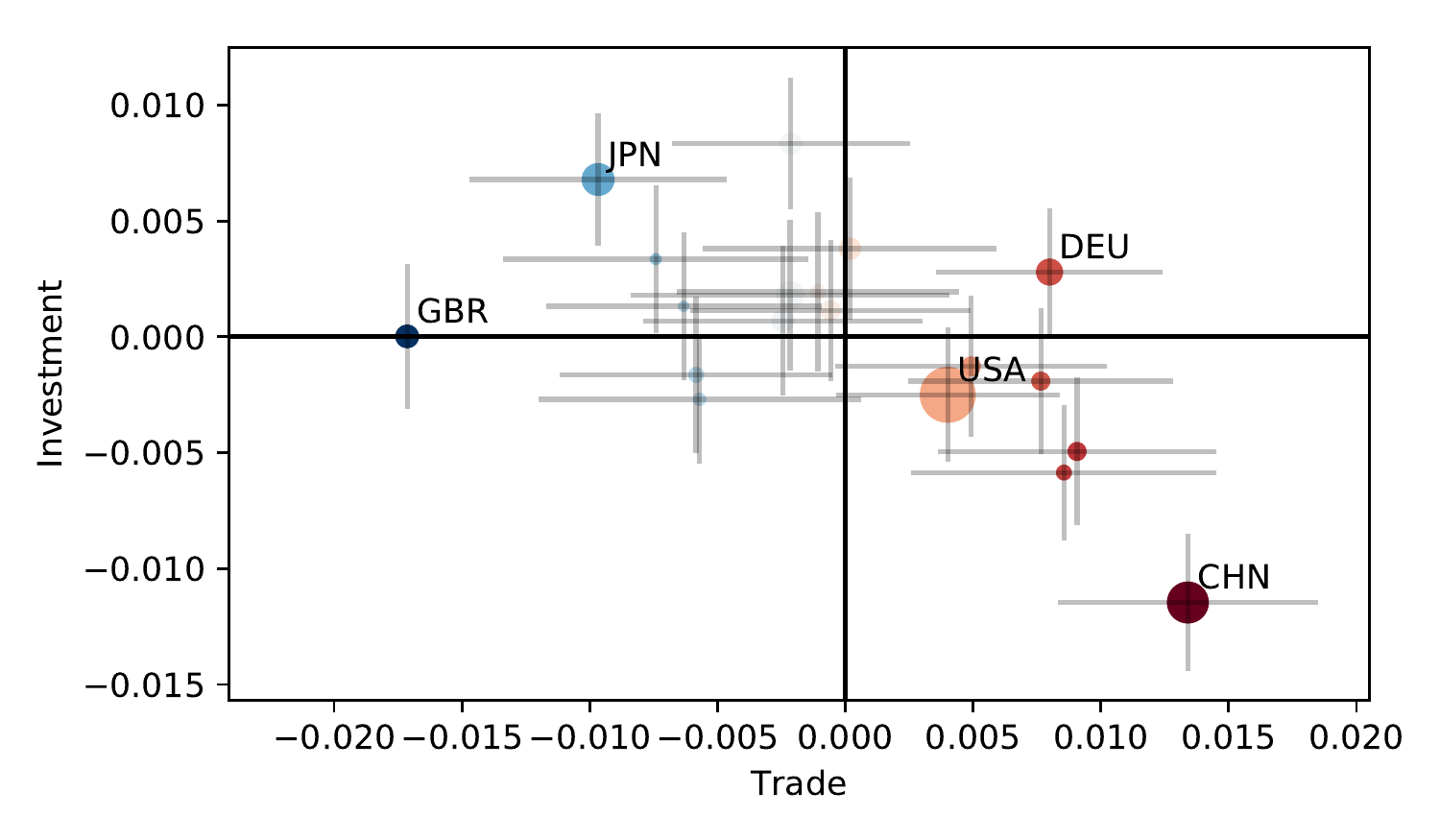}    
    \includegraphics[width=8.5cm,angle=0]{\FigPath/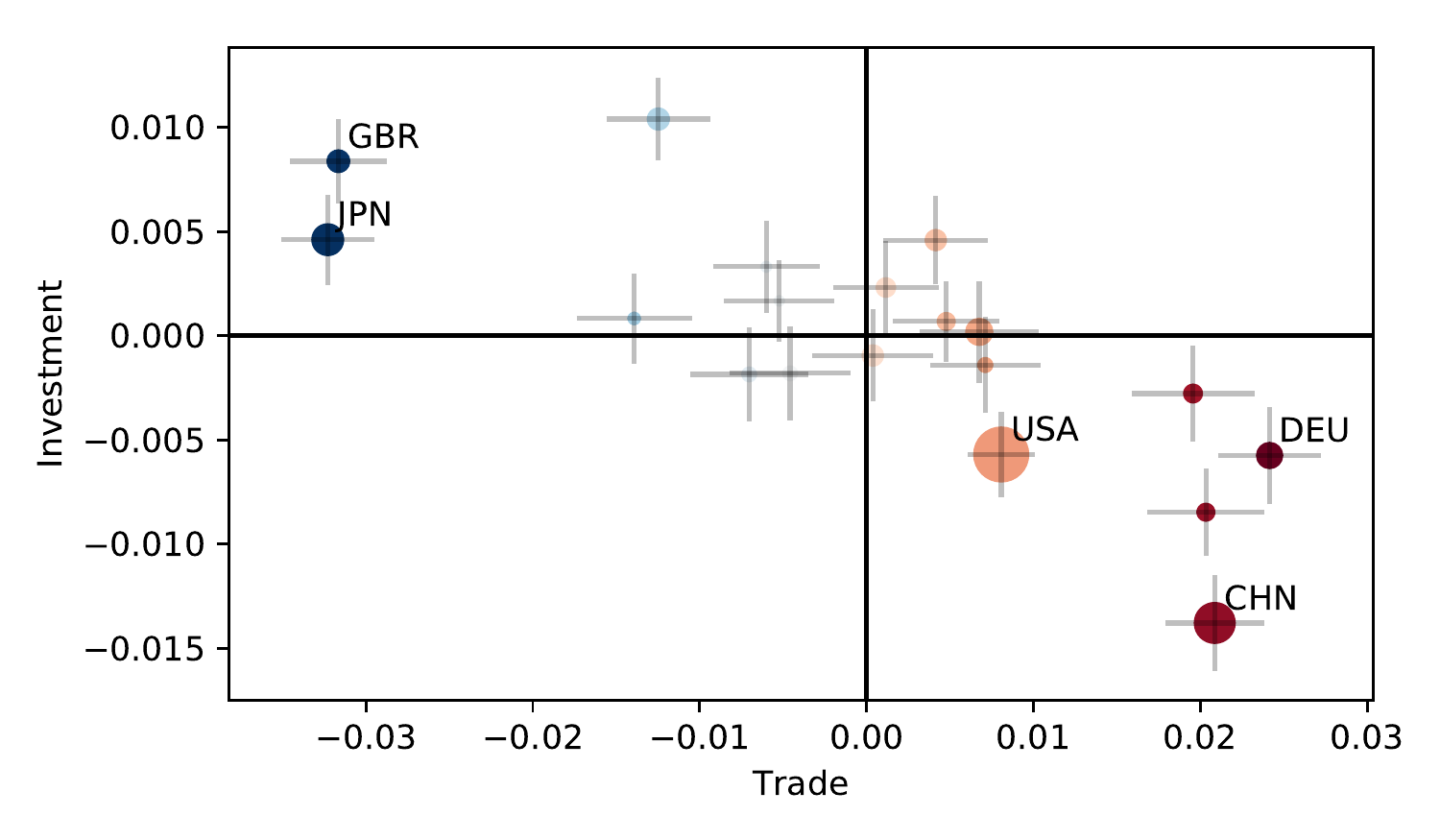}  
      \end{center}
      \caption{ Trade (x-axis) versus  financial ( y-axis) deviations, as obtained by 
      Figure \ref{fig:sys_imp}. The initial shock is characterized  by $\alpha = -0.1, \beta = -0.3$ (left), or $\alpha =-0.3, \beta = -0.5$ (right).  
      Countries belonging to the $G_{20}$ group are shown. Error bars represent the standard error of the mean for the systemic impact. 
      Size of dots is proportional to countries' GDP. 
         \label{fig:residuals} }
\end{figure}

\newpage

\section{Network multipliers predict systemic impact}

Here we show the systemic impact as a function of the an initial shock originated only in the investment (Fig. \ref{fig:gen_imp_in_shock_fin}) or trade  (Fig. \ref{fig:gen_imp_in_shock_trade})  layer. We also show the comparison between actual and predicted systemic impacts, obtained by means of the network multipliers presented in the main text.   
 These Figures correspond to Figs.~4 and 5 of the main text, with different values of $\alpha$ and $\beta$.
 
\begin{figure}[tbp]
  \begin{center}  
    \includegraphics[width=8.5cm,angle=0]{\FigPath/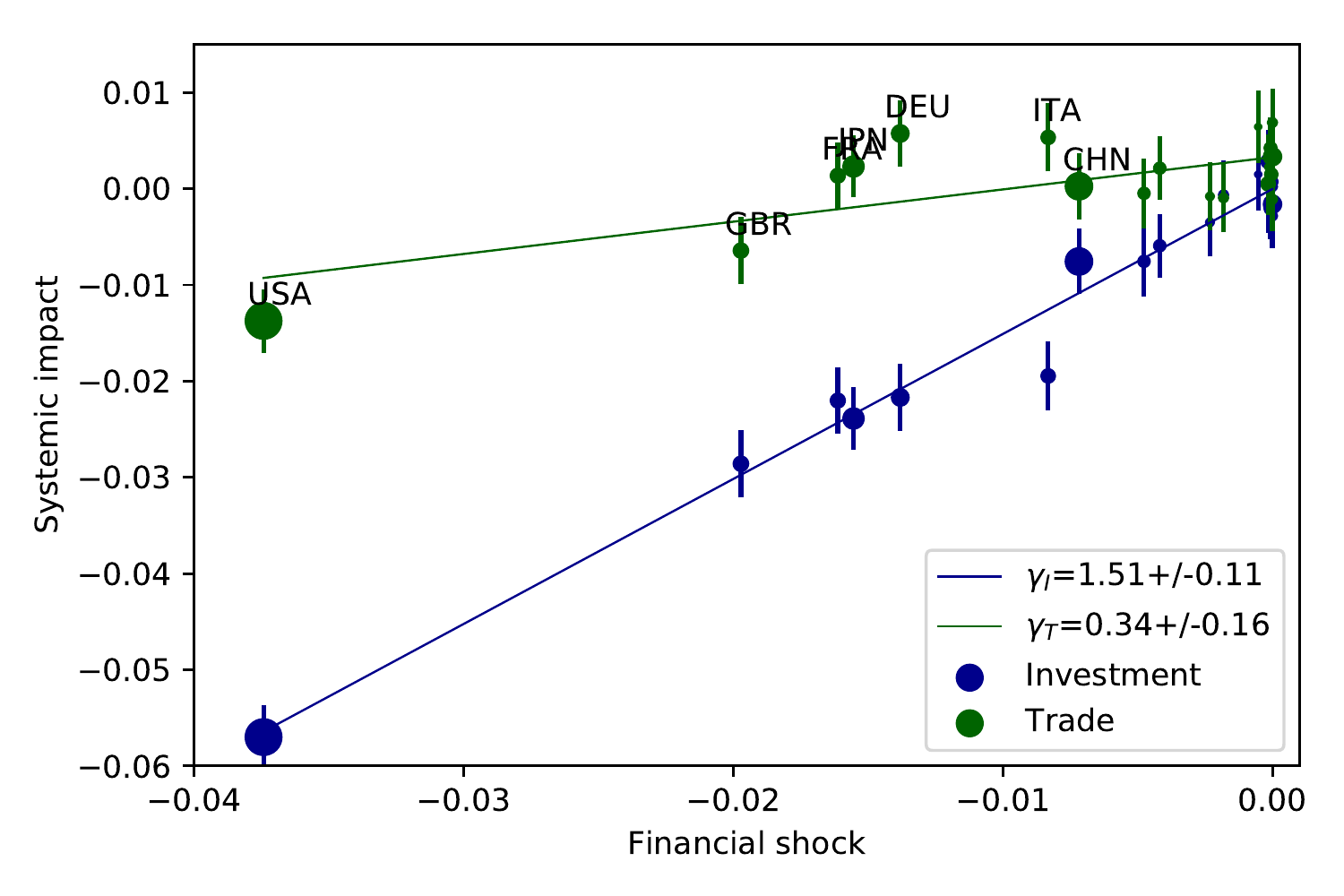}    
    \includegraphics[width=8.5cm,angle=0]{\FigPath/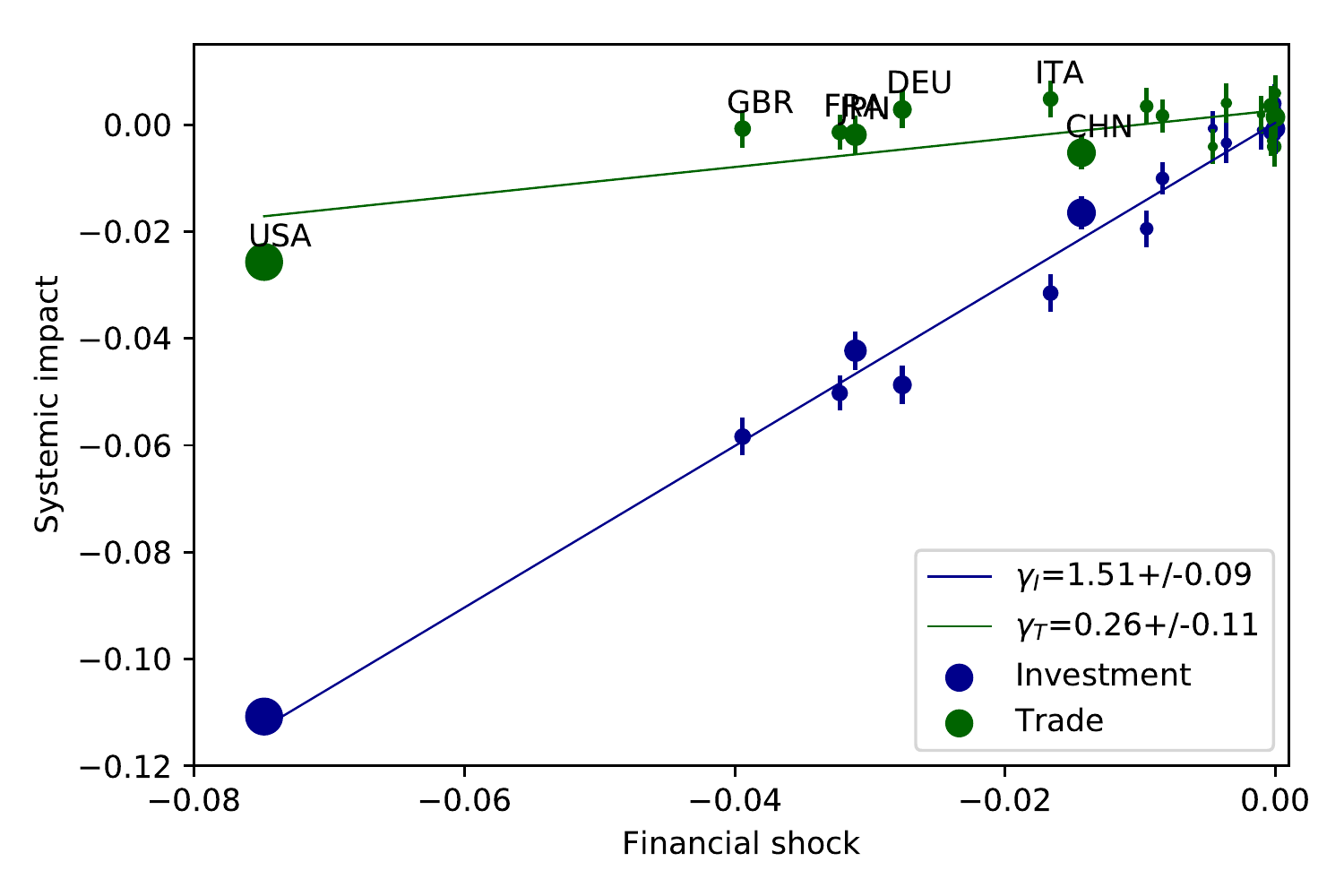}    
      \end{center}
      \caption{ 
     Systemic impact on global trade $\mathcal{S}_i^T$ and investment $\mathcal{S}_i^I$, as a function of the an initial shock $\mathcal{I}_i^{\ell}/W_{\ell}$ originated only in the investment layer, for $\beta=-0.2$ (left) or $\beta=-0.4$ (left), for countries belonging to the $G_{20}$ group. Error bars represent the standard error of the mean for  $\mathcal{S}_i$. Regression coefficients $\gamma_{\ell' \rightarrow \ell}$ are plotted with $95\%$ CI. Size of dots is proportional to countries' GDP.  
       \label{fig:gen_imp_in_shock_fin}}    
\end{figure}

\begin{figure}[tbp]
  \begin{center}  
    \includegraphics[width=8.5cm,angle=0]{\FigPath/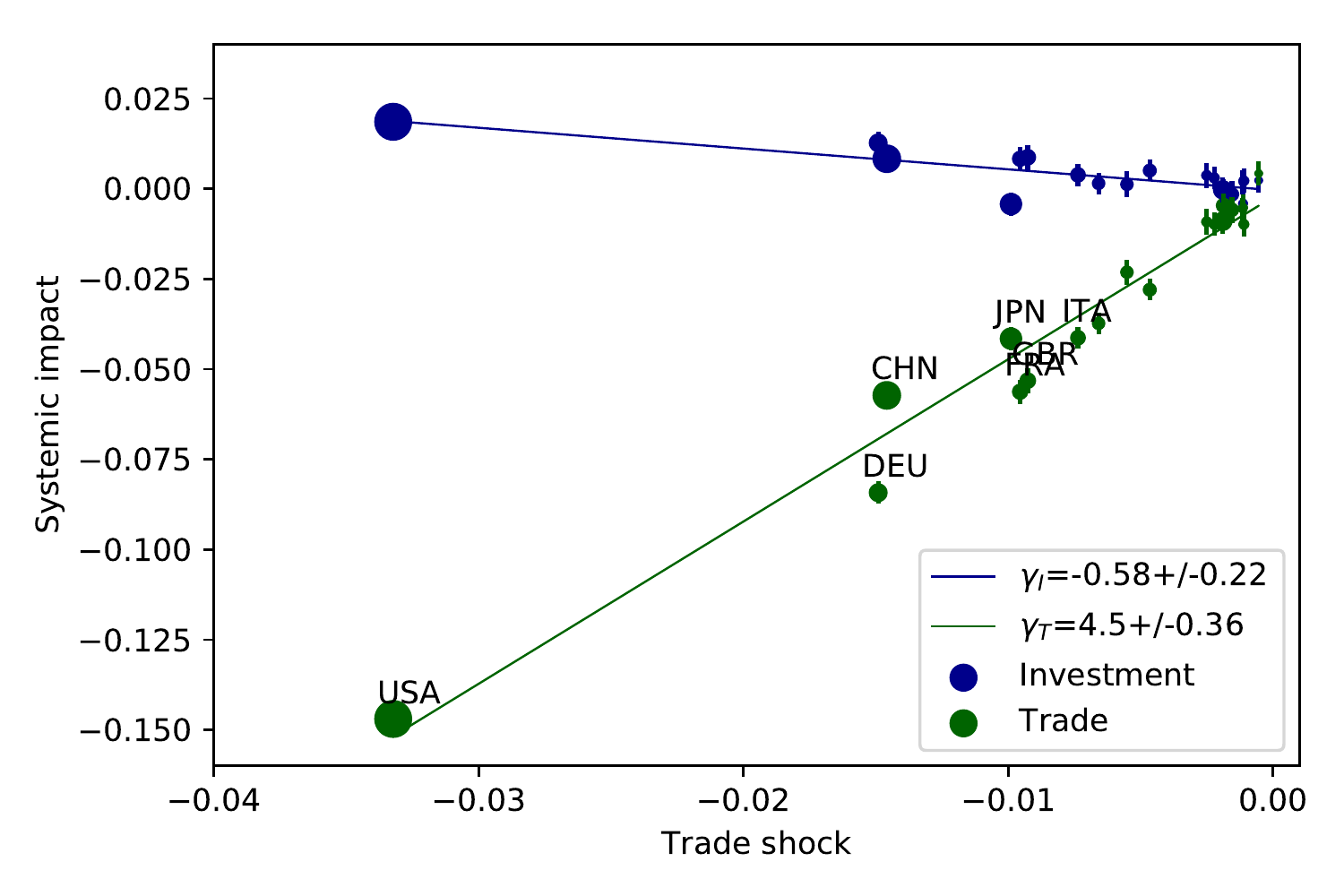}    
    \includegraphics[width=8.5cm,angle=0]{\FigPath/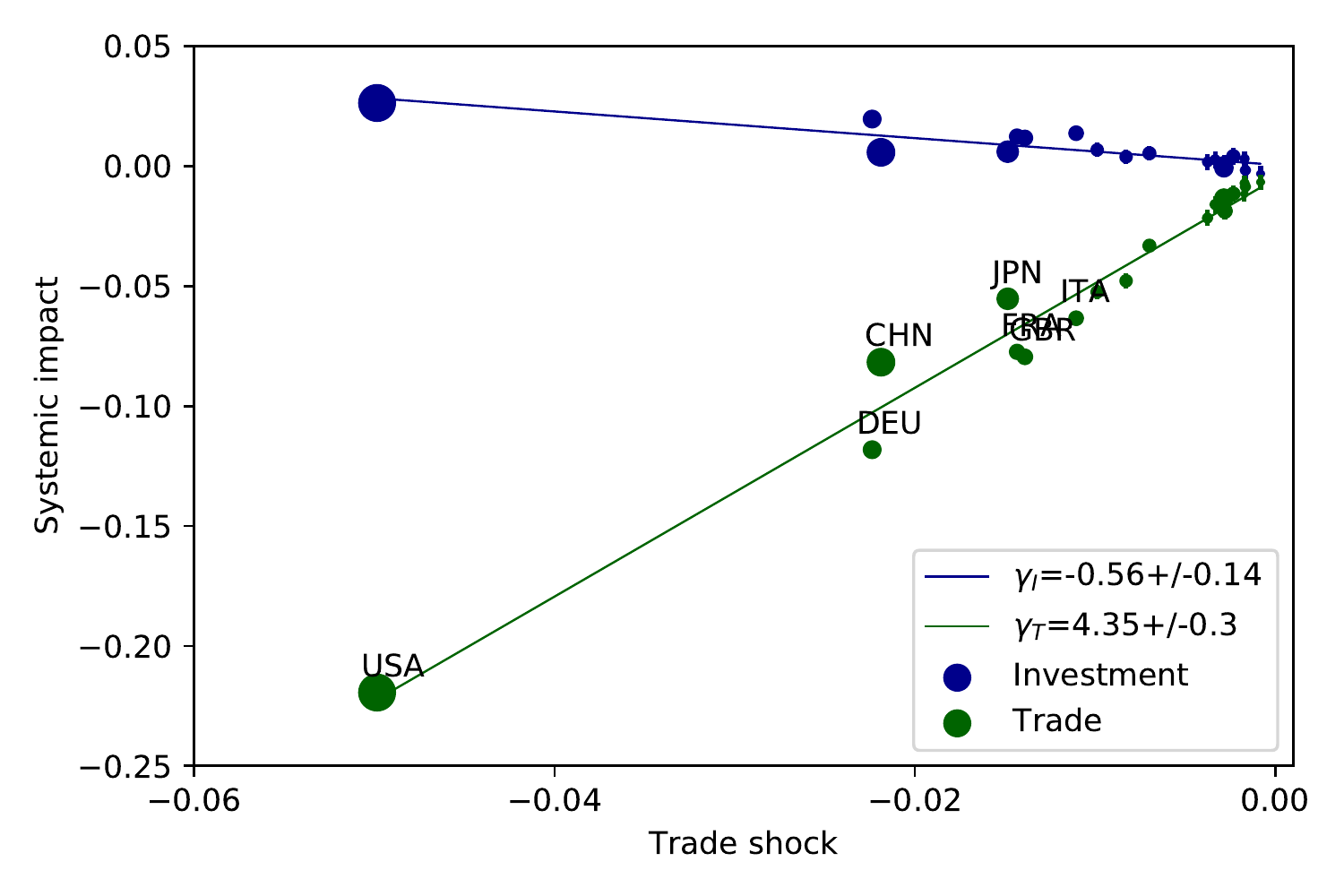}    
      \end{center}
      \caption{ 
     Systemic impact on global trade $\mathcal{S}_i^T$ and investment $\mathcal{S}_i^I$, as a function of the an initial shock $\mathcal{I}_i^{\ell}/W_{\ell}$ originated only in the trade layer, for $\alpha=-0.2$ (left) or $\alpha=-0.3$ (right), for countries belonging to the $G_{20}$ group. Error bars represent the standard error of the mean for  $\mathcal{S}_i$. Regression coefficients $\gamma_{\ell' \rightarrow \ell}$ are plotted with $95\%$ CI. Size of dots is proportional to countries' GDP.  
       \label{fig:gen_imp_in_shock_trade}}     
\end{figure}

\begin{figure}[tb]
  \begin{center}  
      \includegraphics[width=8.5cm,angle=0]{\FigPath/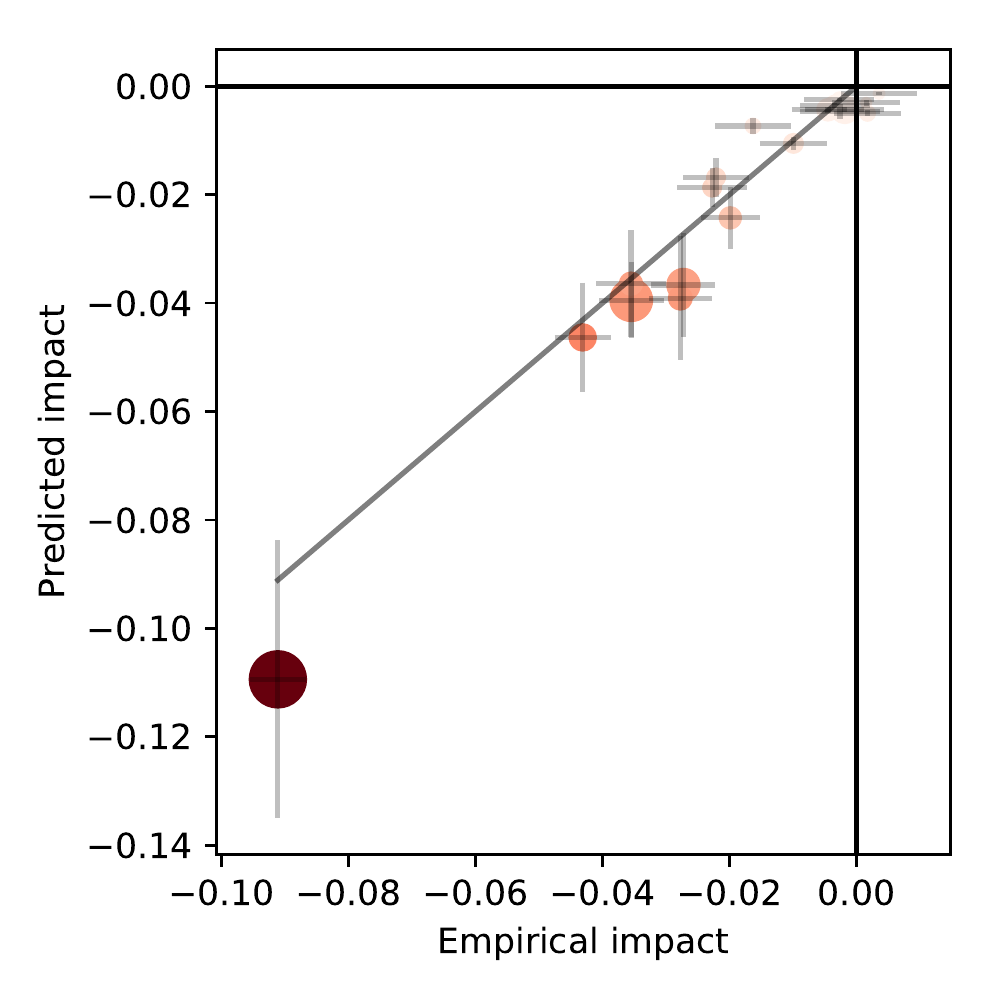}
      \includegraphics[width=8.5cm,angle=0]{\FigPath/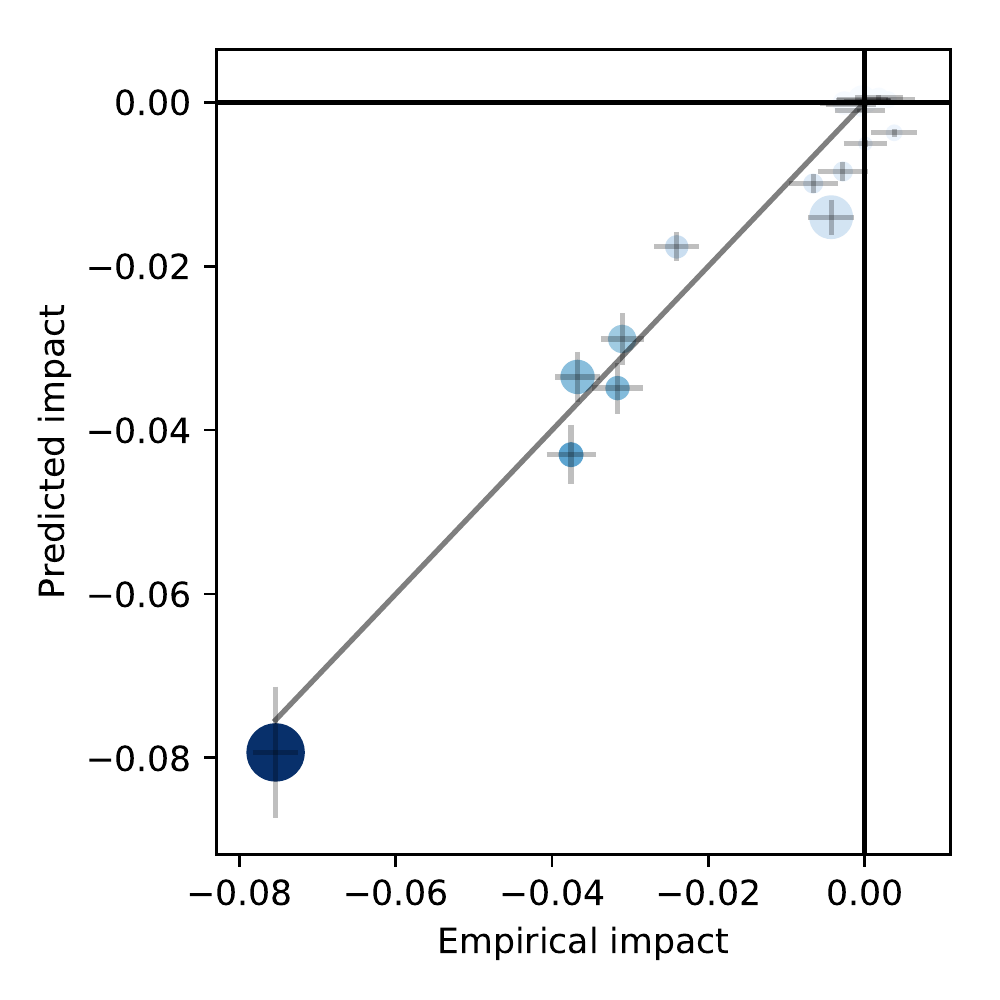}
      \includegraphics[width=8.5cm,angle=0]{\FigPath/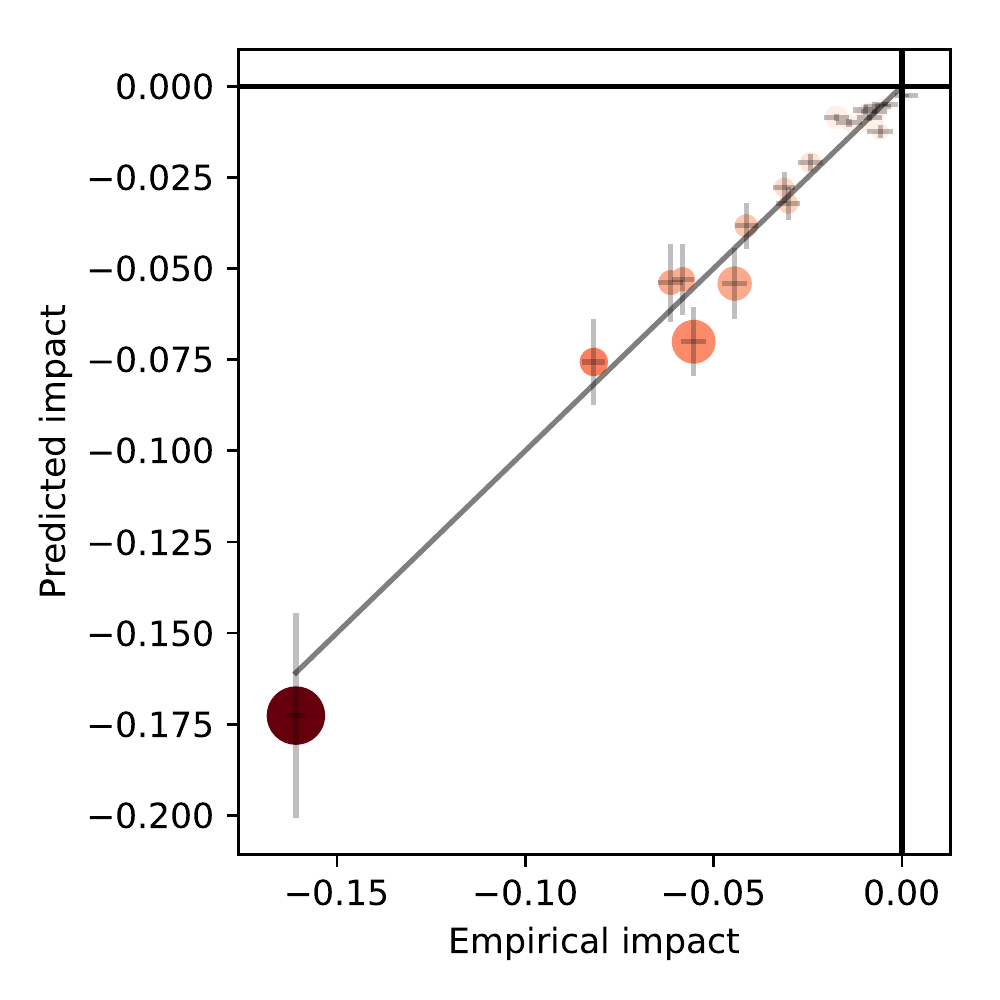}
      \includegraphics[width=8.5cm,angle=0]{\FigPath/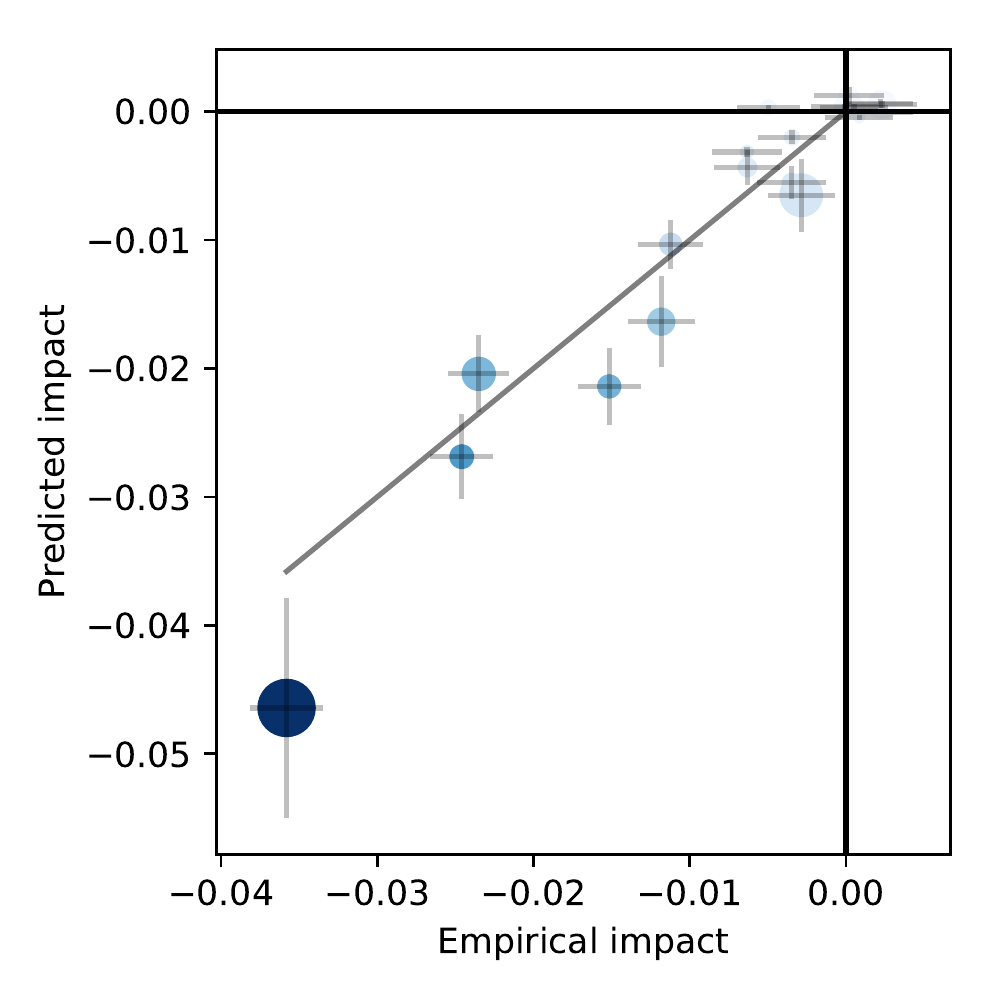}
\end{center}
      \caption{ Expected versus actual systemic impact on trade (left) and investment (right) of each country $i$ belonging to the $G_{20}$ group, originated by an initial shock with $\alpha = -0.1, \beta = -0.3$ (top), or   with $\alpha = -0.2, \beta = -0.2$ (bottom). 
      The size of dots is proportional to their GDP, color proportional to $\mathcal{S}_i^{\ell}$ (red for $\ell=T$, blue for $\ell=I$). 
      Uncertainties are represented by grey crosses.  \label{fig:pred} }
\end{figure}

\end{widetext}


\end{document}